\documentclass[a4paper,11pt]{article}
\pdfoutput=1 
\usepackage{jheppub} 
\usepackage[T1]{fontenc} 
\usepackage{mathrsfs}
\usepackage{graphicx, amsthm, multirow,soul}
\usepackage[citecolor=blue]{hyperref}
\usepackage[all]{hypcap}
\usepackage{url}
\usepackage{color}
\usepackage{subfigure}
\usepackage{slashed}
\usepackage[dvipsnames]{xcolor}
\usepackage{multicol, blindtext}
\usepackage[section]{placeins}
\usepackage{lipsum}
\newcommand{\equaref}[1]{Eq.~(\ref{#1})}
\newcommand{\equasref}[2]{Eqs.~(\ref{#1})~and~(\ref{#2})}

\newcommand{\figref}[1]{Fig.~\ref{#1}}

\newcommand{\secref}[1]{Section~\ref{#1}}

\newcommand{\appref}[1]{Appendix~\ref{#1}}

\usepackage{amsmath}
\usepackage{latexsym}
\usepackage{stackengine,scalerel}
\newcommand{\bq}{\begin{eqnarray}}
\newcommand{\nq}{\end{eqnarray}}
\newcommand{\no}{\nonumber}
\newcommand{\tr}{\text{tr}}

\newcommand{\cM}{\mathcal{M}}
\newcommand{\bfk}{\mathbf{k}}
\newcommand{\bfq}{\mathbf{q}}
\newcommand{\bfx}{\mathbf{x}}
\newcommand{\bfr}{\mathbf{r}}
\newcommand{\bfp}{\mathbf{p}}
\newcommand{\Imag}{\text{Im}}
\newcommand{\Real}{\text{Re}}
\newcommand{\gm}{\gamma}
\DeclareGraphicsExtensions{.pdf,.jpg,.png}
\usepackage{epstopdf}
\preprint{\hspace{10.5mm}  IPPP/18/65  \hspace{10.5mm}FERMILAB-PUB-18-329-T }

\title{\boldmath Leptogenesis via Varying Weinberg Operator: \\   the Closed-Time-Path Approach}

\author[a]{Jessica Turner}
\author[b]{and Ye-Ling Zhou}

\affiliation[a]{Theoretical Physics Department, Fermi National Accelerator Laboratory, P.O. Box 500, Batavia, IL 60510, USA.}
\affiliation[b]{Institute for Particle Physics Phenomenology, Department of Physics,
Durham University, Durham DH1 3LE, United Kingdom.}

\emailAdd{jturner@fnal.gov}
\emailAdd{ye-ling.zhou@durham.ac.uk}

\abstract{
In this work we provide a detailed study of the CP violating phase transition (CPPT) which is a new mechanism proposed to produce a baryon asymmetry. This mechanism exploits the Weinberg operator whose coefficient is dynamically realised from the vacuum expectation values (VEVs) of new scalars. In the specific case of the first order phase transition, the scalar VEVs vary in the bubble wall which separates the two phases. This results in a spacetime varying coefficient for the Weinberg operator. The interference of two Weinberg operators at different spacetime points generates a CP asymmetry between lepton and anti-lepton production/annihilation processes, which eventually results in an asymmetry between baryon and anti-baryon number densities in the early Universe. 
We present the calculation of the lepton asymmetry, 
based on non-equilibrium quantum field theory methods, in full. We  
consider the  influence of the bubble wall characteristics and the 
impact of thermal effects on the lepton asymmetry and draw a comparison between the CPPT mechanism and electroweak
baryogenesis.
}

\begin{document} 
\maketitle
\flushbottom

\newpage
\section{Introduction}\label{sec:intro}
It remains a mystery and fundamental open question how our visible Universe to be matter dominated. 
The abundance of  matter over anti-matter is approximately given by \cite{PDG}
\bq
5.8  \times 10^{-10} <\eta_B \equiv \frac{n_B-n_{\overline{B}}}{n_\gamma} <6.6 \times 10^{-10} ~(95\%~\text{CL})\,,
\nq
where $n_B$,  $n_{\overline{B}}$ and $n_{\gamma}$ are the number densities of baryons, anti-baryons and photons respectively. 
Although the Standard Model (SM)  provides baryon-number-violating and lepton-number-violating (LNV) processes while preserving the $B-L$ number, it does 
not contain sufficient sources of CP-violation or a sufficiently fast departure from thermal equilibrium to generate the observed asymmetry. 
Baryogenesis via leptogenesis, as first proposed by Fukugita and Yanagida \cite{leptogenesis}, is one of the most widely studied explanations of the origin of the matter-antimatter asymmetry in the early universe. In their mechanism, they proposed that a lepton asymmetry is generated above the  electroweak (EW) scale through the
CP-asymmetric decays of heavy Majorana neutrinos. The lepton asymmetry is subsequently partially converted into a baryon asymmetry via $(B-L)$-preserving weak sphaleron processes \cite{Khlebnikov:1988sr}.

A particularly strong motivation for leptogenesis is its connection with small but non-zero neutrino masses. In order to understand the origin of neutrino masses, most theoretical studies support that neutrinos are Majorana in nature and their masses are obtained from the well-known dimension-five Weinberg operator \cite{Weinberg}
\bq
\mathcal{L}_\text{W}=\frac{\lambda_{\alpha\beta}}{\Lambda} \ell_{\alpha L} H C \ell_{\beta L} H + \frac{\lambda_{\alpha\beta}^*}{\Lambda} \overline{\ell_{\alpha L}} H^* C \overline{\ell_{\beta L}} H^*\,,
\label{eq:Weinberg} 
\nq
where $\lambda_{\alpha\beta}=\lambda_{\beta\alpha}$ are effective Yukawa couplings with flavour indices $\alpha,\beta=e,\mu,\tau$, $C$ is the charge conjugation matrix and $\Lambda$ is the scale of the new physics responsible for neutrino masses. It is an obvious but important point to note this operator violates lepton number. After EW symmetry breaking, the Higgs acquires a vacuum expectation value (VEV), $\langle H \rangle = v_H/\sqrt{2}$ with $v_H = 246$ GeV, and neutrinos gain Majorana masses. The $(\alpha,\beta)$ entry of the neutrino mass matrix, $m_\nu$, given by
\bq 
(m_{\nu})_{\alpha\beta} = \lambda_{\alpha\beta} \frac{v_H^2}{\Lambda} \,. 
\nq 
If we assume  a dimensionless coefficient $\lambda \sim \mathcal{O}(1)$, an $\mathcal{O}(0.1)$ eV scale neutrino mass is naturally obtained for $\Lambda \sim \mathcal{O}(10^{14})$ GeV. It is worth stressing, the Weinberg operator violates lepton number and $B-L$ symmetry. At tree-level, this dimension-five  operator may be ultraviolet (UV) completed  through the introduction of  fermionic singlets \cite{Mohapatra:1979ia,GellMann:1980vs,Yanagida:1979as,Minkowski:1977sc}, scalar triplets \cite{Schechter:1980gr,Mohapatra:1980yp,Magg:1980ut,Lazarides:1980nt,Wetterich:1981bx} or fermionic triplets \cite{Foot:1988aq,Ma:1998dn}  which are known as the type-I, II and III see-saw mechanisms respectively. Alternatively, it is possible (Majorana) neutrino masses are generated via loop induced processes \cite{Zee:1980ai,Babu:1988ki,Ma:2006km}.  Moreover, there have been proposals that neutrinos masses derive from effective operators
with dimension greater than five \cite{Chen:2006hn,Gogoladze:2008wz} or from large extra-dimensions \cite{Grossman:1999ra,ArkaniHamed:1998vp}. 

For decades various models involving new symmetries have been proposed to address  neutrino properties. Many models related to the neutrino mass generation assume a $U(1)_{B-L}$ symmetry \cite{Minkowski:1977sc, GellMann:1980vs, Yanagida:1979as, Schechter:1981cv} at sufficiently high energy scale. The tiny neutrino masses are obtained after the breaking of this symmetry. 
In a series of flavour models, the observed pattern of lepton mixing is generated by the breaking of some underlying flavour symmetries. A large number of symmetry groups have been considered, from continuous ones such as $U(1)$ \cite{Froggatt:1978nt}, $SO(3)$ \cite{Berger:2009aa, King:2018fke}, $SU(3)$ \cite{Alonso:2013nca}, and also the  
discrete case $Z_n$ \cite{Grimus:2004hf,Zhou:2012ds}, $A_4$ \cite{Ma:2001dn,Altarelli:2005yp,Altarelli:2005yx}, 
$S_4$ \cite{Mohapatra:2003tw,Lam:2008rs}
$\Delta(27)$ \cite{deMedeirosVarzielas:2005qg,deMedeirosVarzielas:2006fc}, $\Delta(48)$ \cite{Ding:2013nsa,Ding:2014hva}, etc. For a comprehensive review see e.g., Refs \cite{Altarelli:2010gt,King:2013eh,King:2014nza}. 
An important motivation for the current and next-generation neutrino experiments is the measurement of leptonic CP violation.
These experimental endeavours have triggered 
many theoretical studies of CP violation in the lepton sector. In particular, what is the nature of CP violation? 
Is CP symmetry broken spontaneously \cite{Lee:1973iz,Branco:1979pv} or explicitly? If spontaneous symmetry breaking occurs is it geometric in nature \cite{deMedeirosVarzielas:2011zw,Branco:1983tn} or compatible with flavour symmetries \cite{Feruglio:2012cw,Holthausen:2012dk}? 

The implications for leptogenesis, in the context of many of these neutrino mass generation mechanisms,  have been explored in a great number of
works.  
In order to generate a lepton asymmetry above the electroweak scale, all such mechanisms must satisfy Sakharov's three conditions \cite{Sakharov:1967dj}: $B-L$ violation; and C/CP violation; and out-of-equilibrium dynamics\footnote{This statement assumes  CPT is a conserved symmetry. There are theories which propose CPT-violation as a means of baryogenesis \cite{Bertolami:1996cq}.}. 
There are indirect means of testing these conditions in the lepton sector.

\textbf{Lepton number violation} is inextricably linked to the Majorana 
nature of neutrinos. This property of neutrinos will be tested by the undergoing 
\cite{Capelli:2005jf,BRUGNERA:2014ava,Albert:2014afa,Asakura:2014lma,Ardito:2005ar}
and future planned \cite{Sibley:2014nda,DavidLorcafortheNEXT:2014fga,Fritts:2013mwa,Xu:2015dfa,Nova:2013ata,Ishihara:2012pwa} neutrinoless double beta decay experiments. Leptonic mixing and {\bf CP violation} may be constrained from the complementarity between reactor neutrino experiments,  such as Daya Bay \cite{An:2012eh}, RENO \cite{Ahn:2012nd} and Double Chooz  \cite{Ardellier:2006mn}, and long base-line accelerator 
experiments such as T2K  \cite{T2K} and NO$\nu$A \cite{Ayres:2004js} which have shown a  slight statistical  preference for maximally CP violation with $\delta\sim 3\pi/2$. 
The next generation of neutrino oscillation experiments such as  DUNE \cite{DUNE} and T2HK \cite{T2HK} will be able to make precision measurement of this phase.

There are a number of distinct types of leptogenesis and the energy scale of each mechanism depends upon the nature of  the departure from thermal equilibrium.  As previously mentioned, in the original paper \cite{leptogenesis} the out-of-equilibrium dynamics are  provided by the CP-asymmetric decays of Majorana neutrinos. 
The lower bound on the temperature, and therefore heavy Majorana neutrino mass scale, needed to successfully generate sufficient lepton asymmetry is above $10^{9}$ GeV \cite{Davidson:2002qv}\footnote{This \emph{Davidson-Ibarra} bound has several caevats: \textbf{(i)} flavour effects are negligible,  \textbf{(ii)}
the heavy Majorana mass spectrum is hierarchical and  \textbf{(iii)} the lightest heavy Majorana neutrino dominantly contributes to the lepton asymmetry.}. 
Thermal leptogenesis may be lowered to the TeV scale if the heavy Majorana neutrinos are near degenerate in mass as this causes a resonant enhancement of the CP asymmetry \cite{Pilaftsis:1997jf,Pilaftsis:2003gt,Pilaftsis:2005rv,Dev:2017wwc}. 
 In addition, the out-of-equilibrium dynamics may be provided by means other than the decays of heavy see-saw mediators. In the Akhmedov-Rubakov-Smirnov (ARS) mechanism \cite{Akhmedov:1998qx}, this is realised by the smallness of the Yukawa coupling $y_\text{D}$ between $\ell$ and heavy Majorana neutrinos. 
For alternative mechanisms involving heavy seesaw mediators, see e.g., \cite{Kim:2016xyi, Hambye:2016sby}.

In \cite{Pascoli:2016gkf}, we proposed a novel mechanism of leptogenesis which proceeds via a time-varying Weinberg operator which is present during a phase transition (PT). 
As explained therein and shall be discussed in depth later, this mechanism satisfies the three Sakharov conditions as follows: 
\begin{itemize}
\item
The Weinberg operator violates lepton in addition to $B-L$ number. 

\item
The Weinberg operator is out of thermal equilibrium at temperatures $T<10^{13}$ GeV.  

\item
We assume a CP-violating PT (CPPT), which results in a time-varying coefficient in the Weinberg operator.
\end{itemize}
Using the Weinberg operator to fulfil the Sakharov condition is not new and has been considered in, e.g., \cite{Kusenko:2014uta, Hamada:2015xva, Takahashi:2015ula}.
Through the combination of the three Sakharov conditions we  arrive at an out-of-equilibrium spacetime-varying CP-violating Weinberg operator. While the Weinberg operator induces lepton and anti-lepton production/annihilation processes in the thermal plasma, the interference of the varying Weinberg operator at two different spacetime points generates a CP asymmetry between them. Eventually, a net lepton asymmetry is generated after the PT. As the lepton asymmetry is increased by the temperature, we found the minimal temperature for 
successful baryogenesis to be approximately $T_\text{CPPT} \sim 10^{11}$ GeV. CPPT  is crucially reliant upon  the scale of the PT to be below the scale at which the  Weinberg operator decouples from the theory, $T_{\text{CPPT}}<\Lambda$. Otherwise, heavy particles in the UV sector have not decoupled and may wash out the lepton asymmetry generated by the PT.  
Therefore, a key 
difference between our leptogenesis mechanisms and all others is that the New Physics responsible for light neutrino mass generation has been
integrated out \emph{before} the CP-violating processes become active, and consequently CPPT is  independent of the specific neutrino mass model.
Moreover, this implies in CPPT the CP-violation scale is \emph{below} the neutrino mass generation scale. 

The application of a PT in the context of leptogenesis has not been well studied in the literature. Beyond our work,
authors in
\cite{Pilaftsis:2008qt,Shuve:2017jgj}
 explored the effects of a phase transition on the baryon asymmetry generated via out-of-equilibrium
decays. In particular, they discussed  the scenario where  the parent particle responsible for
baryogenesis obtains its mass via spontaneous symmetry breaking and phase transitions in the early universe
 gives rise to a time-dependent mass of the right-handed neutrino. 
Another scenario, in the framework of the type-I seesaw with an $U(1)_{B-L}$ symmetry,  has recently been discussed in \cite{Long:2017rdo}. 
They suggested an asymmetry between the heavy Majorana neutrino ($N$) and its CP-conjugate
 is initially generated  in front of the bubble wall, where $U(1)_{B-L}$ is preserved and $N$ is massless. After the heavy Majorana neutrinos diffuse into the $U(1)_{B-L}$-breaking bubble and acquire masses, the $N$-$\overline{N}$ asymmetry produces  a lepton asymmetry through the decay of heavy Majorana neutrinos. 
Our mechanism distinctly differs from these models as the lepton asymmetry is generated after the physics responsible for neutrino  masses has been integrated out. However, the three mechanisms share the common feature that they proceed via a cosmological phase transition (PT).  
 
The main purpose of this work is to provide a detailed analysis of the mechanism proposed in \cite{Pascoli:2016gkf}. In 
\secref{sec:VWO} we motivate and discuss the mechanism in full generality. We follow in \secref{sec:KBE} with a brief  review of the Closed-Time-Path (CTP) formalism used to  obtain the lepton asymmetry via the Kadanoff-Baym (KB) equation. The CTP approach together with KB equation is a powerful tool to calculate non-equilibrium thermal processes \cite{Schwinger:1960qe, Keldysh:1964ud, KB}. It has seen wide and successful application in the EW baryogenesis (EWBG) \cite{Riotto:1998zb,Carena:2000id,Carena:2002ss,Lee:2004we}, leptogenesis via heavy Majorana neutrino decays \cite{Anisimov:2010aq,Anisimov:2010dk,Frossard:2012pc,Garbrecht:2013urw,Garbrecht:2013gd}, resonant leptogenesis
\cite{Garbrecht:2011aw,Dev:2014laa,Dev:2014aa} and ARS mechanism \cite{Drewes:2012ma,Drewes:2016gmt}. Using this approach, we need not consider individual processes separately, but instead include all processes in the CP-violating self energy corrections. Moreover, unlike semi-classical calculations, memory effects are properly accounted for in this formalism. In \secref{sec:calclep}, we analyse in detail how the generated lepton asymmetry is influenced by the bubble wall properties and thermal effects of the leptons and the Higgs. 
We assume a single scalar PT to simplify the discussion. 
Our numerical analysis is provided in \secref{sec:numerical}. Finally, we summarise and  make concluding remarks in \secref{sec:final}. From Appendices~\ref{sec:EEVprofle} to \ref{sec:ME}, we list examples of the EEV profiles, extend our discussion to the multi scalar PT and list the details of the element matrix calculation. We specifically highlight the main differences between our mechanism and EWBG in Appendix~\ref{sec:EWBG} and discuss of the influence the oscillation effect in the varying Weinberg operator in Appendix~\ref{sec:oscillating}. We refer to Ref.~\cite{Pascoli:2018cqk} for a semi-classical approximation of this mechanism. 

%%%%%%%%%%%%%%%%%%%%%%%%%%%%%%%
%%%%    VARYING WEINBERG OPERATOR %%%%%%%
%%%%%%%%%%%%%%%%%%%%%%%%%%%%%%%

\section{Varying Weinberg Operator}\label{sec:VWO}
In the Standard Model (SM), tiny neutrino masses may be explained by introducing higher-dimensional operators. The simplest operator is the dimension-five Weinberg operator of \equaref{eq:Weinberg} which violates lepton number and generates Majorana masses for neutrinos.
In many New Physics models, the coefficient of the Weinberg operator $\lambda_{\alpha\beta}$ in \equaref{eq:Weinberg} is not a fundamental parameter; rather is dynamically realised after some scalars acquire VEVs. In this section, we will discuss how to achieve a varying Weinberg operator and introduce the mechanism of leptogenesis via the varying Weinberg operator.

%%%%%%%%%%%%%%%%%%%%%%%%%%%%%%%
%%%%    MOTIVATIONS %%%%%%%
%%%%%%%%%%%%%%%%%%%%%%%%%%%%%%%
\subsection{Motivations of the Varying Weinberg Operator}\label{sec:motivation}

We begin with two UV-complete toy models to illustrate how the varying Weinberg operator may be obtained.  These two models differ 
from each other in how the  scalar VEV contributes to the neutrino mass. For simplicity, we assume a single scalar, $\phi$.
 The corresponding Lagrangian terms in these two models (referred as Model I and Model II) are respectively given by 
\begin{eqnarray}
\hspace{-5mm}\mathcal{L}_{\text{I}} \ &=& \sum_{\alpha, I, J} y_{\alpha I} \overline{N_{IR}} H \ell_{\alpha L} - \frac{1}{2} \kappa^*_{IJ} \phi^* N_{IR} C N_{JR} - \frac{1}{2} (M^0_N)^*_{IJ} N_{IR} C N_{JR}  + \text{h.c.}\,, \no\\
\hspace{-5mm}\mathcal{L}_{\text{II}} &=& \sum_{\alpha, a, b, I, J} y^0_{\alpha I} \overline{N_{I R}} H \ell_{\alpha L} + x_{\alpha a} \overline{\Psi_{a R}} H \ell_{\alpha L}  + z_{a I} \phi \overline{N_{IR}} \Psi_{a L}  \no \\
&&+ (M_\Psi)_{ab} \overline{\Psi_{bR}} \Psi_{aL} - \frac{1}{2} (M_N)^*_{IJ} N_{IR} C N_{JR} + \text{h.c.}\,, 
\end{eqnarray}
where $\alpha=e,\mu,\tau$ is the charged lepton flavour and $N$ the heavy Majorana neutrino with index  $I$, $\Psi$ a heavy vector-like fermions with index $a$, and $y_{\alpha I}$, $x_{\alpha a}$ and $z_{a I}$ are dimensionless constant coefficients. 

In these two models, $\phi$ plays a different role in the light neutrino mass generation as can be  clearly seen if we assume the scalar gets a VEV, $v_\phi$, before the decoupling of any heavy particles. 
In Model I, $\phi$ contributes to the Majorana mass term for the heavy neutrino, $N$. After $\phi$ acquires a VEV, the mass matrix for $N$ is given by $M_N=M^0_N + \kappa v_\phi$. 
In Model II,  $\phi$ contributes to the Dirac mass term between light neutrinos and heavy neutrinos. 
By assuming the $\Psi$ mass is sufficiently heavy, the decoupling of $\Psi$ results in a higher dimensional operator between $\ell$ and $N$, $(xM^{-1}_\Psi z)_{\alpha I} \phi \overline{N_{I R}} H \ell_{\alpha L}$, where $(M_\Psi)_{ab} = M_a \delta_{ab}$. After $\phi$ gets a VEV, we arrive at an effective Yukawa coupling $y = y^0 + xM^{-1}_\Psi z v_\phi$. After the decoupling of heavy neutrinos, we obtain the Weinberg operator with the coefficient of the Weinberg operator given by $\lambda = y M^{-1}_N y^T$ in both models.

Now let us assume the decoupling of heavy new states occurs before the PT. After this decoupling, one can effectively express the Weinberg operator with the coefficient $\lambda$ given by 
\bq
\lambda = y (M^0_N + \kappa \phi)^{-1} y^T = y (M^0_N)^{-1} y^T - \left[y (M^0_N)^{-1} \kappa (M^0_N)^{-1} y^T \right] \phi + \cdots,
\nq
for Model I, and 
\bq
\lambda &=& (y^0+x M^{-1}_\Psi z \phi) M^{-1}_N (y^0+x M^{-1}_\Psi z \phi)^T \no\\
&=& y^0 M^{-1}_N y^{0\, T} + \left[ x M^{-1}_\Psi z M^{-1}_N y^{0\,T} + y^0 M^{-1}_N z^T  (M^{T}_\Psi)^{-1} x^T \right] \phi + \cdots,
\nq
for Model II. Before the PT, $\phi$ is zero valued, so the coefficient $\lambda$ is identical to $\lambda^0 = [y (M_N^0)^{-1} y^T$ in Model I or $\lambda^0 = [y^0 M_N^{-1} y^{0\,T}]^*$ in Model II, which is different from the coefficient after the PT,  $\lambda = y M_N^{-1} y^T$. In other words, we encounter a varying Weinberg operator during the PT which is a consequence of the PT occurring after  heavy particle decoupling.

It is straightforward to generalise the above discussion to a PT with multiple scalars. Assuming the PT happens after the heavy particles decouple,  the coefficient of the Weinberg operator in the most generic case is written
\bq
\lambda_{\alpha\beta} = \lambda^0_{\alpha\beta} + \sum_{i=1}^n \lambda^{i}_{\alpha\beta}   \frac{ \phi_i }{v_{\phi_i}} + \sum_{i,j=1}^n \lambda^{ij}_{\alpha\beta}  \frac{ \phi_i }{v_{\phi_i}} \frac{ \phi_j }{ v_{\phi_j}} + \cdots \,, 
\nq
where $n$ represents the number of  scalars,  $\lambda^0$, $\lambda^i$, $\lambda^{ij}$, ... are a set of constant coupling matrices in the flavour space with $\alpha,\beta = e,\mu,\tau$ are flavour indices. These couplings  are determined by the details of neutrino models, in particular by the assumed new symmetries. It is worth noting that although we have introduced heavy neutrinos, based on type-I seesaw, to obtain the Weinberg operator in the toy models; the UV structure is really irrelevant for us to obtain the varying Weinberg operator. Replacing the heavy neutrinos of the type-I seesaw with heavy particles from type-II, III seesaws or radiative models, one can derive similar spacetime-dependent couplings, $\lambda_{\alpha\beta}$, after all heavy particles decouple. 

The breaking of the symmetry may be achieved by the scalars acquiring non-zero VEVs, $\langle \phi_i \rangle = v_{\phi_i}$, and in turn the coefficient of the Weinberg operator is dynamically realised, $\lambda_{\alpha\beta} = \lambda^0_{\alpha\beta} + \sum_i \lambda^i_{\alpha\beta} + \sum_{i,j} \lambda^{ij}_{\alpha\beta} + \cdots$. To generate CP violation in $m_\nu$, there must be some phases which cannot be reabsorbed by rephasing in
$\lambda^0_{\alpha\beta}$, $\lambda^i_{\alpha\beta}$, $\lambda^{ij}_{\alpha\beta}, \cdots$. These phases may arise explicitly or spontaneously and both possibilities have been studied  extensively in many  models. 

In a thermodynamical system, the ensemble expectation value (EEV) of an operator $\mathcal{A}$  is described by $\langle \mathcal{A} \rangle = \text{Tr} (\rho\mathcal{A})$, where $\rho$ is the density matrix of the statistical ensemble. 
In the early Universe at high temperature, the EEVs of $\phi_i$ is dependent on the structure of the scalar potential at finite temperature. 
In the very early Universe, the vacuum is in the  symmetric phase, $\langle \phi_i \rangle = 0$. As the Universe expands and cools, 
the vacuum at $\langle \phi_i \rangle = 0$ becomes metastable and the PT proceeds to the true and asymmetric vacuum $\langle \phi_i \rangle = v_{\phi_i}$. 

\begin{figure}[t]
\centering
\includegraphics[width=0.8\textwidth]{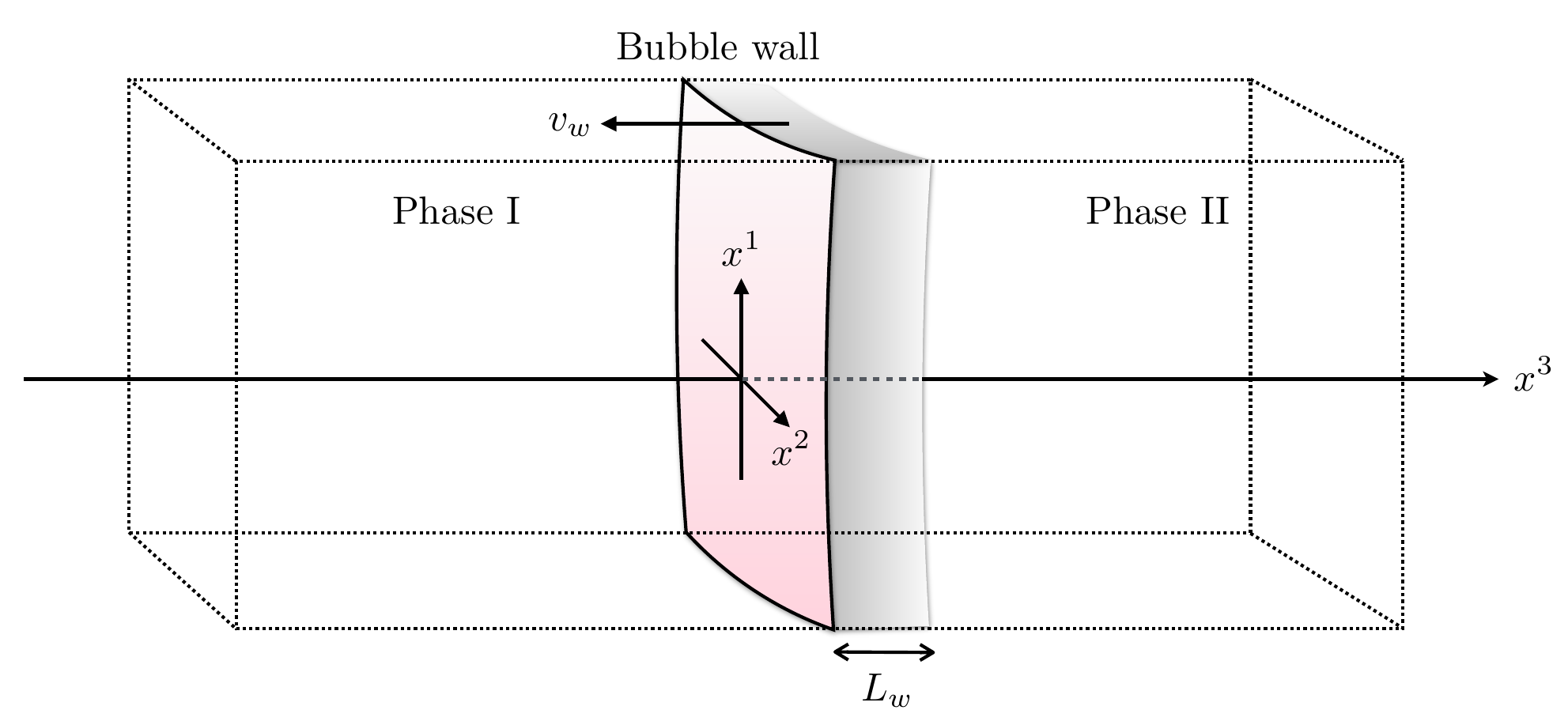}
\caption{\label{fig:bubble}The true vacuum ($\langle\phi_i\rangle \neq v_{\phi_i}$) expanding over the false vacuum ($\langle\phi_i\rangle = 0$). The width of the bubble wall and its expansion velocity are denoted as $L_w$  and $v_w$ respectively. }
\end{figure}

In the following, we limit our discussion to a first-order PT, which is not qualitatively crucial for the mechanism to be successful but
allows for straightforward interpretation and can simplify the calculation as we shall discuss later.
During this PT, 
bubbles of asymmetric phase (labelled as Phase II) nucleate, via thermal tunnelling \cite{Linde:1977mm,Linde:1978px}, and  expand in the symmetric phase (labelled as Phase I). 
 We characterised the width of the bubble wall as $L_w$ and the expansion velocity as $v_w$ in the $-x^3$ direction, as shown in \figref{fig:bubble}. In the bubble wall, the averaged value of $\lambda$ is a time- and space-dependent value, which we denote as 
\bq
\lambda_{\alpha\beta}(x) \equiv |\lambda_{\alpha\beta}(x)|e^{i \phi_{\alpha\beta}(x)} \,.
\nq 

%%%%%%%%%%%%%%%%%%%%%%%%%%%%%%%
%%%%    MECHANISM %%%%%%%
%%%%%%%%%%%%%%%%%%%%%%%%%%%%%%%
\subsection{The Mechanism of Leptogenesis}\label{sec:mechanism}

The Weinberg operator may trigger the following lepton number violating (LNV) processes:
\bq
& H^* H^* \leftrightarrow \ell \ell\,,\quad
\overline{\ell} H^* \leftrightarrow \ell H\,,\quad
\overline{\ell} H^* H^* \leftrightarrow \ell \,,\no\\
&\overline{\ell} \leftrightarrow \ell H H\,,\quad 
H^* \leftrightarrow \ell \ell H\,,\quad
0 \leftrightarrow \ell \ell H H
\label{eq:LNV_processes}
\nq
and their CP conjugate processes. 
Of the processes shown in  \equaref{eq:LNV_processes}, the right pointing arrow denotes lepton production in the thermal plasma while the 
left pointing arrow indicates lepton annihilation. The CP conjugation processes lead to the anti-lepton production and annihilation.
Given a fixed spatial point during the PT, the coefficient of the Weinberg operator changes with time. Therefore, Weinberg operators at different times may interact with each other, and through their interference may produce a lepton asymmetry. However, a departure from thermal equilibrium is necessary and in order to understand how this is achieved, we may consider the  Hubble expansion rate: 
\begin{itemize}
\item The Hubble expansion scale $H_u$, represents how fast the early Universe expands and is given by
\bq
H_u \approx \sqrt{g_*} \frac{T^2}{M_{\text{pl}}} \,,
\nq
where $M_{\text{pl}}=1.22\times10^{19}$ GeV is the Planck mass and  $g_*$ is the effective number of degrees of freedom contributing to the energy density in the early Universe. In the Standard Model, $g_*=106.75$. 

\item The Weinberg operator reaction scale $\Gamma_\text{W}$ characterises how fast the LNV processes occur. We assume this mechanism occurs at temperatures much higher than the EW scale, such that the Higgs has not yet acquired a non-zero  VEV and  are thermally distributed. The rate of these processes is approximately 
\begin{eqnarray}
\Gamma_\text{W} \approx \frac{3}{4\pi^3}\frac{\lambda^2}{\Lambda^2} T^3 \approx \frac{3}{4\pi^3}\frac{m_\nu^2}{v_H^4} T^3\,,
\end{eqnarray}
where we have parametrised $\lambda$ by the neutrino mass $m_\nu$ ($m_{\nu}=\lambda v^2_H/\Lambda$). 
\end{itemize}

For temperature $T<10^{13}$ GeV, the Weinberg operator reaction scale $\Gamma_{\text{W}}$ is smaller than the Hubble expansion rate $H_u$. 
As a consequence of the smallness of $\Gamma_{\text{W}}$, any LNV processes resulted from the Weinberg operator are out of thermal equilibrium. On the other hand, the \textbf{washout effects} triggered by the Weinberg operator, are not efficient because $\Gamma_{\text{W}}$ is so small. In conventional methods of leptogenesis, the see-saw mediators may participate in interactions which washout the lepton asymmetry. In this  mechanism the scale of the PT, triggering the leptogenesis, occurs below the scale of neutrino mass generation and therefore CPPT does not suffer from this  type of washout.

One may  wonder if the scalar, $\phi$, modifies the out-of-equilibrium dynamics and contributes to washout processes via the operator $\frac{\lambda^i}{\Lambda} \frac{\phi^i}{v_{\phi_i}} (LH)^2$. The reaction rate of this operator, $\Gamma_{\phi_i}$, depends on the mass and VEV of 
$\phi$. Naively, we may assume  they are of the same order as the temperature $T$. In this case, $\Gamma_{\phi_i} \ll \Gamma_{\text{W}}$ and as a consequence of  the phase space suppression implies these interactions may be safely neglected. From these remarks, it is clear that the interactions  of the Weinberg operator themselves are out of thermal equilibrium and the PT is not necessary to satisfy Sakharov's second condition.
A possible exception to this conclusion is the scenario of the 
 $\phi_i$ mass, $m_{\phi_i}$, being much larger than the temperature $T$.  If this is the case, then $\phi_i$ will decay very quickly after the PT, with decay rate
\begin{eqnarray}
\Gamma_{\phi^i} \sim \frac{1}{8\left(4\pi\right)^5} \frac{\text{tr}[\lambda^i \lambda^{i*}]}{\Lambda^2} \frac{m_{\phi_i}^5}{v_{\phi_i}^2}\,. 
\end{eqnarray} 
This reaction rate would be much larger than $\gamma_{\text{W}}$ or even larger than the Hubble expansion rate and  a net lepton asymmetry may be produced through the decay of $\phi_i$. However, there will still be no  washout as the backreaction of  $\phi_i$ decays are suppressed. This particular possibility will not be  considered further in this   paper. 

There are other scales in this problem. Although  they shall not ultimately determine if this mechanism works, they will play an important quantitative role in the final calculation of the lepton asymmetry: 
\begin{itemize}
\item The damping rate of the Higgs and leptons $\gamma_{H,\ell}$. These damping rates are mainly determined by the SM interactions, $\gamma_{H,\ell} \sim 0.1 \,T$ \cite{bellac}.  These rates are related to the inverse mean free paths $1/L_H$ and $1/L_\ell$ and represent how fast these particles decouple from the LNV interactions. 
\item The dynamics of the PT. In particular, the bubble wall scale  (i.e., the inverse wall thickness $1/L_w$) and the wall velocity, $v_w$, in the case of first-order PT. The parametric regime of these parameters indicates how fast the bubble wall sweeps over a certain region, and how quickly the  false vacuum is replaced by the true one.
\end{itemize}

These two important properties of the bubbles  will influence both the lepton asymmetry and 
the cosmological imprint CPPT leaves in the Universe. 
There are two parametric regimes the bubble wall characteristic  may assume:

\begin{itemize}
\item The nonadiabatic ``thin wall'' regime: $L_w \ll L_{H,\ell}$. The wall is thinner than the mean free paths of the relevant particles. We shall mainly focus on this case  because it allows us to integrate out the full lepton asymmetry without considering the detailed properties of the bubble wall as shown below. 

\item The adiabatic ``thick wall'' regime: $L_w \gg L_{H,\ell}$. The thick wall case has been widely used in the EW phase transition, where the Higgs wall thickness is constrained by the Higgs mass and EW scale. In the thick wall case, the lepton asymmetry is dependent upon how the $\phi$ VEV evolves in the wall.  A brief discussion of this scenario  can be found in \appref{sec:PT2}.
\end{itemize}

Both the thickness of the bubble wall and its velocity are model-dependent features determined from the scalar potential of $\phi$ and thermal corrections from the SM particles in the thermal plasma \cite{Anderson:1991zb,Dine:1992wr,Moore:2000jw}. The bubble wall velocity is crucially dependent upon the pressure difference across the  wall and the friction induced 
 on the wall by the plasma.  The friction is calculated 
from a set of Boltzmann equations coupled to the motion of the scalar field  and this
effect is related to the deviation from equilibrium in the plasma \cite{Moore:1995si,Bodeker:2017cim,Bodeker:2009qy}. 
In CPPT,  $\phi_i$  couples only to the leptons and the Higgs thus we find it a reasonable assumption
that the bubble walls of CPPT are fast moving. 
For simplicity we assume a thin wall and relegate more model-dependent studies to future work.

%%%%%%%%%%%%%%%%%%%%%%%%%%%%%%%
%%%%    KB EQUATIONS   				%%%%%%%
%%%%%%%%%%%%%%%%%%%%%%%%%%%%%%%
\section{Kadanoff-Baym Equation in the Closed-Time-Path Approach} \label{sec:KBE}

%%%%%%%%%%%%%%%%%%%%%%%%%%%%%%%
%%%%    CTP  						%%%%%%%
%%%%%%%%%%%%%%%%%%%%%%%%%%%%%%%
\subsection{Closed-Time-Path Formalism} 
Before we discuss the relevant details of the Closed-Time-Path formalism, we shall motivate its use through a brief discussion 
of
the semi-classical approach,  an alternative method,  of calculating the time  evolution of the particle number density for a given process. 
These semi-classical kinetic equations are typically derived from \emph{Liouville's} equation which 
states that the probability distribution function ($f$) of a system of particles does not change along any trajectory in phase space.
Liouville's equation details the evolution of an $n$-particle system and hence the probability distribution function  in 6$n$-dimensional phase space (three position and three momentum coordinates are needed to describe each particle). Using the Poisson bracket, this equation may be written in the following manner
\bq\label{eq:poisson}
 \frac{\partial f}{\partial t}  = \{H,f\} \quad \text{where} \quad \{A,B\} =  \frac{\partial A}{\partial \mathbf{r}_{i}} .\frac{\partial B}{\partial \mathbf{p}_{i}} - \frac{\partial A}{\partial \mathbf{p}_{i}} .\frac{\partial B}{\partial \mathbf{r}_{i}},
\nq
where $H$ is the Hamiltonian of the system,  $\mathbf{r}$ and $\mathbf{p}$ are position and momentum respectively. For generic systems, the distribution function is dependent on a very large number of variables ($\sim 10^{23}$) and solving \equaref{eq:poisson} quickly becomes intractable.  The first step in simplifying these equations is to apply the Bogoliubov-Born-Green-Kirkwood-Yvon (BBGKY) hierarchy \cite{BBGKY1,BBGKY2,BBGKY3} which allows the  $n$-distribution function to be written as a function of the $n+1$ distribution function (essentially $f_{1}=\mathcal{F}(f_2), f_{2}=\mathcal{F}(f_3),... $). These sets of recursive equations are just as difficult as \equaref{eq:poisson} to solve. However, in the limiting case where  the system of particles may be considered as a dilute gas these equations can be truncated such that the time evolution of the system is represented by the one-particle distribution function\footnote{In the dilute gas approximation the timescale of the collisions ($t_{C_{i}}$) is much smaller than the timescale of the particles propagating between collisions ($t_{prop}$) i.e.\ $t_{C_{i}}\ll t_{prop}$}   ($f_1$)
\bq\label{eq:BE}
\frac{\partial f_{1}}{\partial t} = \{H_{1},f_{1}\} + \left(\frac{\partial f_{1}}{\partial t}\right)_{\text{coll}},
\nq
where the third term of \equaref{eq:BE} is the collision integral and accounts for scattering between particles\footnote{The semi-classical Boltzmann equation of \eqref{eq:BE} is a standard result of kinetic theory and some standard steps have been skipped.}. Such scatterings are calculated using S-matrix elements in the usual \emph{in-out} formalism  at zero temperature. From the Lehmann-Symanzik-Zimmermann (LSZ) reduction formula,  S-matrix elements are expressed  in terms of correlation functions of fields which are asymptotically free of each other; in a dilute gas this approximation is reasonable given that the timescale of collisions between particles is significantly shorter than the timescale of particle propagation and thus the in-coming and out-going are asymptotically free states. \\
\begin{figure}[t]
\centering
\includegraphics[width=.95\textwidth]{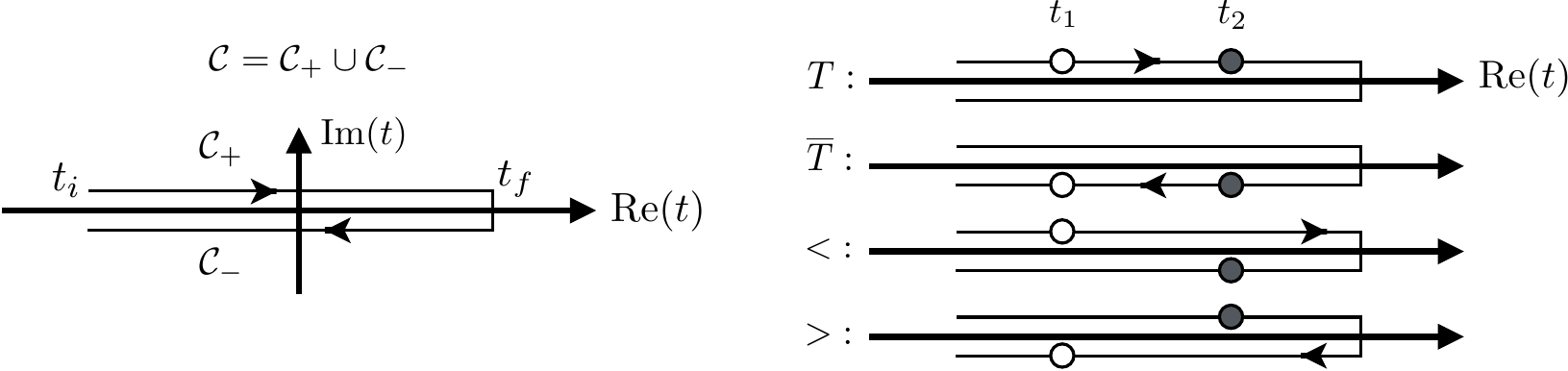}
\caption{\label{fig:CTP}Left panel: the CTP time contour. The time runs from an initial time $t_i$ to some final time $t_f$ and then returns to $t_i$. Right panel: time ordered ($T$), anti-time ordered ($\overline{T}$), $t_1\to t_2$ ($<$) and $t_2 \to t_1$ ($>$) paths defined in the CTP propagators.}
\end{figure} 
One may question the validity of such a treatment  in the finite temperature and density environment of the Early Universe.
 Therefore, representing the  system as a dilute gas may not be fully descriptive given that the timescale of particle propagation may not be significantly different from the timescale of the collisions; in such a scenario each subsequent particle
collision may be influenced by a history of collisions 
and therefore the system becomes non-Markovian in nature. To capture such memory effects amounts to going beyond the \emph{in-out} formalism, based on zero-temperature S-matrix elements as discussed previously, to using the \emph{in-in} formalism. This formalism may also be known as  the Real-Time, Closed-Time Path (CTP) and  Schwinger-Keldysh formalism \cite{Schwinger:1960qe, Keldysh:1964ud}. Regardless of the nomenclature, the benefit of using such an approach allows the assumption of asymptotically free states to be removed.

Such non-equilibrium dynamics requires the specification of an initial state. This corresponds to a special choice of the time contour, the Closed-Time-Path $\mathcal{C}=\mathcal{C_+}\bigcup \mathcal{C_-}$ with $\mathcal{C_+}$ evolving from an initial time $t_i$ to some final time $t_f$ and then $\mathcal{C_-}$ evolves backwards, as shown in the left panel of \figref{fig:CTP}. In the CTP approach, propagators are defined depending upon which contours the spacetime points $x_1$ and $x_2$ are localised. We may simplify the CTP propagators into four propagators: Feynman (time ordered, represented by $T$), Dyson (anti-time ordered, represented by $\overline{T}$), and Wightman (the order from $t_1 \equiv x_1^0$ to $ t_2 \equiv x_2^0$, represented by $<$ and the order from $t_2$ to $t_1$, represented by $>$) propagators, as shown in the right panel of \figref{fig:CTP}. 

For the Higgs ($H$), the propagators in the CTP approach is defined as
\bq
(\Delta^{\mathcal{C}})_{ab}^{}(x_1,x_2) &=& \begin{pmatrix} 
(\Delta^{T})_{ab}^{}(x_1,x_2) & (\Delta^{<})_{ab}^{}(x_1,x_2) \\ 
(\Delta^{>})_{ab}^{}(x_1,x_2) & (\Delta^{\overline{T}})_{ab}^{}(x_1,x_2) \\ \end{pmatrix},
\nq
where the Feynman, Dyson and Wightman propagators of the Higgs $\Delta^{T}$, $\Delta^{\overline{T}}$ and $\Delta^{<,>}$ are defined to be
\bq
(\Delta^{T})_{ab}^{}(x_1,x_2) &=& \langle T[H_a(x_1) H_b^*(x_2)] \rangle \,, \text{ for } t_1,t_2 \in \mathcal{C_+}\,,\no\\
(\Delta^{\overline{T}})_{ab}^{}(x_1,x_2) &=& \langle \overline{T}[H_a(x_1) H_b^*(x_2)] \rangle \,, \text{ for } t_1,t_2 \in \mathcal{C_-}  \,, \no\\
(\Delta^{<})_{ab}^{}(x_1,x_2) &=& \langle H_b^*(x_2) H_a(x_1) \rangle \,,\hspace{5mm} \text{ for } t_1 \in \mathcal{C_+}, ~ t_2 \in \mathcal{C_-}\,, \no\\
(\Delta^{>})_{ab}^{}(x_1,x_2) &=& \langle H_a(x_1) H_b^*(x_2) \rangle \,, \hspace{5mm} \text{ for } t_1 \in \mathcal{C_-}, ~ t_2 \in \mathcal{C_+}\,, 
 \label{eq:scalar_propagator}
\nq
respectively. 
In non-equilibrium environments, the system is dependent upon both the relative and average coordinates which are defined by 
 $r=x_1-x_2$ and  $x=(x_1+x_2)/2$ respectively. 
We perform a Wigner transformation to the relative coordinate in the following manner
\bq
\Delta_{k} (x) = \int d^4 r e^{i k\cdot r} \Delta(x+r/2, x-r/2) \,.
\nq 
A general solution for the tree-level propagator is given by
\bq
\Delta^<_{q}(x) &=& 2\pi \delta(q^2) \Big\{ \vartheta(q^0) f_{H, \bfq} (x) + \vartheta(-q^0) [1+ f_{H^*, -\bfq}(x) ] \Big\} \,, \no\\
\Delta^>_{q}(x) &=& 2\pi \delta(q^2) \Big\{ \vartheta(q^0) [1+f_{H,\bfq} (x)] + \vartheta(-q^0) f_{H^*,-\bfq}(x) \Big\} \,,
\label{eq:Higgs_propagator}
\nq
where $f_{H, \bfq} (x)$ and $f_{H^*, \bfq} (x)$ are distribution densities of $H$ and $H^*$, given by the expectation values $\langle a^\dag a\rangle$ and $\langle b^\dag b\rangle$ of free particle and antiparticle mode operators respectively with energy momentum $q^\mu \equiv (q^0, \bfq)$ and $q^2=(q^0)^2-\bfq^2$  \cite{Cirigliano:2009yt}. 

The lepton propagator defined along the CTP contour is defined as
\bq
(S^{\mathcal{C}}_{\alpha\beta})_{ab}^{st}(x_1,x_2) &=& \begin{pmatrix} 
(S^{T}_{\alpha\beta})_{ab}^{st}(x_1,x_2) & (S^{<}_{\alpha\beta})_{ab}^{st}(x_1,x_2) \\ 
(S^{>}_{\alpha\beta})_{ab}^{st}(x_1,x_2) & (S^{\overline{T}}_{\alpha\beta})_{ab}^{st}(x_1,x_2) \\ \end{pmatrix},
\nq
with the Feynman, Dyson and Wightman propagators of the lepton $S^{T}$, $S^{\overline{T}}$ and $S^{<,>}$ given by 
\bq
(S^{T}_{\alpha\beta})_{ab}^{st}(x_1,x_2) &=& \langle T[\ell_{\alpha a}^s(x_1) \overline{\ell}_{\beta b}^t (x_2)] \rangle\,, \text{ for } t_1,t_2 \in \mathcal{C_+}\,, \no\\ 
(S^{\overline{T}}_{\alpha\beta})_{ab}^{st}(x_1,x_2) &=& \langle \overline{T}[\ell^s_{\alpha a} (x_1) \overline{\ell}^t_{\beta b}(x_2)] \rangle\,, \text{ for } t_1,t_2 \in \mathcal{C_-}  \,,\no\\
(S^{<}_{\alpha\beta})_{ab}^{st}(x_1,x_2) &=& -\langle \overline{\ell}^t_{\beta b} (x_2) \ell^s_{\alpha a} (x_1) \rangle \,, \hspace{2mm} \text{ for } t_1 \in \mathcal{C_+}, ~ t_2 \in \mathcal{C_-}\,, \no\\
(S^{>}_{\alpha\beta})_{ab}^{st}(x_1,x_2) &=& \langle \ell^s_{\alpha a} (x_1) \overline{\ell}^t_{\beta b} (x_2) \rangle \,, \hspace{5mm} \text{ for } t_1 \in \mathcal{C_-}, ~ t_2 \in \mathcal{C_+}\,,
 \label{eq:fermion_propagator}
\nq
respectively and 
 the minus sign in $ S^<$ derives from the anti-commutation property of fermions. Flavour indices are denoted by 
$\alpha, \beta$ while  EW gauge  and fermion spinor indices are denoted by $a,b$ and $s,t$  respectively. In the following, we will suppress the EW gauge indices and fermion spinor indices  unless they are stated explicitly. 

The tree-level Wigner transformation of the Wightman propagators $S^{<,>}(x_1,x_2)$ is \cite{Cirigliano:2009yt}
\bq
S^<_{k}(x) &=& -2\pi \delta(k^2) P_L k\!\!\!/ P_R \Big\{ + \vartheta(k^0) f_{\ell, \bfk} (x) - \vartheta(-k^0)[1- f_{\overline{\ell}, -\bfk}(x) ] \Big\} \,, \no\\
S^>_{k}(x) &=& -2\pi \delta(k^2) P_L k\!\!\!/ P_R \Big\{ - \vartheta(k^0) [1-f_{\ell,\bfk} (x)] + \vartheta(-k^0) f_{\overline{\ell}, -\bfk}(x) \Big\} \,,
\label{eq:lepton_propagator}
\nq
where $f_{\ell, \bfk} (x)$ and $f_{\overline{\ell}, \bfk} (x)$ are recognised as distributions with energy momentum $k^\mu \equiv (k^0, \bfk)$ at spacetime around $x^\mu$ of lepton and antilepton respectively and $k^2 = (k^0)^2 - \bfk^2$. 
It is useful to define the following propagators for our later discussion, 
\bq
\begin{aligned}
S^{+}(x_1,x_2) &= \frac{1}{2}[S^{<}(x_1,x_2) + S^{>}(x_1,x_2)]\,, \\
S^{H}(x_1,x_2)& = S^{T}(x_1,x_2)-S^{+}(x_1,x_2)\,. 
\end{aligned}
\nq
These propagators satisfy the following CP properties under the CP transformation, 
\bq
S^{<} (x_1, x_2) \to C S^{>}  (x_2^P,x_1^P) C^{-1} \,,\quad &&
S^{>} (x_1, x_2) \to C S^{<}  (x_2^P,x_1^P) C^{-1} \,,\no\\
S^{+} (x_1, x_2) \to C S^{+}  (x_2^P,x_1^P) C^{-1} \,,\quad &&
S^{H} (x_1, x_2) \to C S^{H}  (x_2^P,x_1^P) C^{-1}\,,
\label{eq:propagator_CP}
\nq
where $(x^{P})^{\mu} \equiv (x^0, -\bfx)$ for $x^\mu = (x^0, \bfx)$. 

In thermal equilibrium, the Higgs and leptons satisfy the Bose-Einstein and Fermi-Dirac distributions which are  respectively 
\bq
&&f_{H, \bfq} = f_{H^*, \bfq} = f_{B,|q^0|} \equiv \frac{1}{e^{\beta |q^0|}-1} \,,\nonumber\\
&&f_{\ell, \bfk} \;= f_{\overline{\ell}, \bfk}\;\;\; = f_{F,|k^0|} \equiv \frac{1}{e^{\beta |k^0|}+1} \,.
\label{eq:equilibrium}
\nq
The relevant tree-level Wightman propagators become spacetime-independent and may be rewritten as
\bq
\Delta^{<,>}_{q}&=& 2\pi \delta(q^2) \Big\{ \vartheta(\mp q^0) + f_{B, |q^0|} \Big\} \,,\no\\
S^{<,>}_{k} &=& 2\pi \delta(k^2) \Big\{ \vartheta(\mp k^0) - f_{F, |k^0|} ] \Big\} P_L k\!\!\!/ P_R \,. 
\label{eq:propagator_equilibrium}
\nq
The Kubo-Martin-Schwinger (KMS) relations are automatically satisfied, $\Delta^{>}_{q} = e^{\beta q^0} \Delta^{<}_{q}$, $S^{>}_{k} = - e^{\beta k^0} S^{<}_{k}$.  In the limiting case as $T\to 0$, the statistical factors $f_{B, |q^0|}, f_{F, |k^0|}$ which correspond to the  thermal contributions tend to zero and hence only the $\vartheta$ terms remain. Thus, 
the $\vartheta$ terms correspond to zero temperature contribution.

%%%%%%%%%%%%%%%%%%%%%%%%%%%%%%%
%%%%    KB EQUATIONS   				%%%%%%%
%%%%%%%%%%%%%%%%%%%%%%%%%%%%%%%
\subsection{Kadanoff-Baym Equation}

The key to calculating the lepton asymmetry is the Kadanoff-Baym equation, which is a component of the Schwinger-Dyson equations based on a 2PI effective action \cite{Prokopec:2003pj, Prokopec:2004ic} in the CTP formalism \cite{Schwinger:1960qe, Keldysh:1964ud}. Assuming a time contour $\mathcal{C}$, the Schwinger-Dyson equation for the left-handed lepton propagator $S_{\mathcal{C}}$ is given by
\bq
i \gamma^\mu\frac{\partial}{\partial x_1^\mu} S_{\mathcal{C}}(x_1,x_3) &=& i \delta_{\mathcal{C}}^4(x_1-x_3) + i \int_{\mathcal{C}} d^4 x_2 \Sigma_{\mathcal{C}}(x_1,x_2)  S_{\mathcal{C}}(x_2,x_3) \,,\no\\
i \frac{\partial}{\partial x_3^\mu} S_{\mathcal{C}}(x_1,x_3) \gamma^\mu&=& i \delta_{\mathcal{C}}^4(x_1-x_3) + i \int_{\mathcal{C}} d^4 x_2 S_{\mathcal{C}}(x_1,x_2)  \Sigma_{\mathcal{C}}(x_2,x_3) \,,
\nq
where $\Sigma_{\mathcal{C}}$ is the self-energy correction to the lepton and all the quantities are time-ordered along the path $\mathcal{C}$. 

The Kadanoff-Baym equation is the equation of motion of the Wightman propagators $S^{<,>}$ and is obtained by decomposing the Schwinger-Dyson equation in the CTP formalism. Its exact expression is given by 
\bq
\begin{aligned}\label{eq:KB}
i \gamma^\mu\frac{\partial}{\partial x_1^\mu} S^{<,>}(x_1,x_3) - \int\! d^4x_2\left\{ \Sigma^{H}(x_1,x_2)  S^{<,>}(x_2,x_3) 
- \Sigma^{<,>}(x_1,x_2)  S^{H}(x_2,x_3)\right\} = \mathcal{C}\\
i\gamma^\mu  \frac{\partial}{\partial x_3^\mu} S^{<,>}(x_1,x_3)- \int\! d^4x_2 \left\{ S^{<,>}(x_1,x_2)  \Sigma^{H}(x_2,x_3) -
 S^{H}(x_1,x_2) \Sigma^{<,>}(x_2,x_3)\right\}  = \overline{\mathcal{C}},
\end{aligned}
\nq
with 
\bq
\mathcal{C}&=& \frac{1}{2}\int d^4x_2 \left[\Sigma^{>}(x_1,x_2)  S^{<}(x_2,x_3) - \Sigma^{<}(x_1,x_2)  S^{>}(x_2,x_3) \right]\,,\no\\ 
\overline{\mathcal{C}}&=& \frac{1}{2}\int d^4x_2 \left[S^{<}(x_1,x_2)  \Sigma^{>}(x_2,x_3) - S^{>}(x_1,x_2)  \Sigma^{<}(x_2,x_3) \right]\,.
\nq
In comparison with the original Schwinger-Dyson equation, the self-energy term $\Sigma_{\mathcal{C}}S_{\mathcal{C}}$ has been divided into three parts in the Kadanoff-Baym equation: \textbf{(i)} $\Sigma^HS^{<,>}$ represents the self-energy contribution to $S^{<,>}$; \textbf{(ii)} $\Sigma^{<,>}S^H$ induces broadening of the on-shell dispersion relation and \textbf{(iii)} $\mathcal{C}$ is the collision term, including the CP source term that is used to generate the lepton asymmetry \cite{Prokopec:2003pj}. 

In the non-equilibrium case, using the Wightman propagators in the momentum space in \equaref{eq:lepton_propagator}, one directly derives  
\bq
\text{tr}[\gamma^\mu i S^+_{k}(x)] = 4\pi \delta(k^2) k^\mu [1- \vartheta(k^0) f_{\ell,\bfk}(x) - \vartheta(-k^0) f_{\overline{\ell},-\bfk}(x)].
\nq 
From the above equation, we integrate over  $k^0$ and the temporal and spatial components are respectively given by
\bq
\begin{aligned}
\int \frac{dk^0}{2\pi}\text{tr}[\gamma^0 i S^+_{k}(x)] &= -\big[f_{\ell,\bfk}(x) - f_{\overline{\ell},-\bfk}(x)\big] \,,\\
\int \frac{dk^0}{2\pi}\text{tr}[\vec{\gamma} i S^+_{k}(x)] &= \hat{\bfk} \big[2 - f_{\ell,\bfk}(x) - f_{\overline{\ell},-\bfk}(x)\big] \,,
\end{aligned}
\nq
where $\hat{\bfk}=\bfk/|\bfk|$. 

The total difference between lepton number and anti-lepton number $\Delta N_\ell \equiv N_\ell-N_{\overline{\ell}}$ in a sufficiently large volume $V = \int d^3\bfx_1$ is defined by
\bq
\Delta N_\ell = \int\! \frac{d^3\bfx_1 d^3\bfk}{(2\pi)^3} \big[f_{\ell,\bfk}(x_1) - f_{\overline{\ell},-\bfk}(x_1)\big] 
&=& -\!\int\! \frac{d^3\bfx_1 d^4 k}{(2\pi)^4} \text{tr}[\gamma^0 i S^+_{k}(x_1)] \no\\
&=& -\!\int\! \frac{d^4 x_1 d^4 k}{(2\pi)^4} \text{tr}[\gamma^0 i \frac{\partial}{ \partial x_1^0 }S^+_{k}(x_1)]. 
\label{eq:define_number_asymmetry}
\nq 
Note that 
\bq
\hspace{-5mm}\int \frac{d^4 x_1 d^4 k}{(2\pi)^4} \text{tr}[\gamma^i i \frac{\partial}{ \partial x_1^i }S^+_{k}(x_1)] =
\int \frac{d t_1 d^3\bfk}{(2\pi)^3} 
\int d^3\bfx_1 \frac{\partial}{\partial x_1^i} \hat{\bfk}^i \big[2 - f_{\ell,\bfk}(x_1) - f_{\overline{\ell},-\bfk}(x_1)\big].
\nq 
In the rest frame of the plasma, we chose the boundaries perpendicular to the $x^3$ direction to be far away from the bubble wall, as shown in \figref{fig:bubble}, such that the mean value of $\hat{\bfk}$ is zero on the boundaries. Using Stokes theorem, the above integration vanishes. Therefore, the lepton asymmetry is simplified to 
\bq
\Delta N_\ell = - \int \frac{d^4 x_1 d^4 k}{(2\pi)^4} \text{tr}[\gamma^\mu i \frac{\partial}{ \partial x_1^\mu }S^+_{k}(x_1)] \,.
\nq 
The lepton asymmetry can be calculated from the Kadanoff-Baym equation. We recall from \equaref{eq:KB} and consider the limit $x_3\to x_1$: 
\bq
i \frac{\partial}{\partial x_1^\mu} \text{tr}\Big[\gamma^\mu S^{+}(x_1,x_1)\Big] &=& \text{tr}\Big[\gamma^\mu i \frac{\partial}{\partial x_1^\mu} S^{+}(x_1,x_3) + i \frac{\partial}{\partial x_3^\mu} S^{+}(x_1,x_3)\gamma^\mu\Big]\Big|_{x_3=x_1},
\nq
where the right-hand side (RHS) of the above may be rewritten as 
\bq
 \int d^4x_2 \Big\{ \text{tr}\Big[\Sigma^{H}(x_1,x_2)  S^{+}(x_2,x_1) - \Sigma^{+}(x_1,x_2)  S^{H}(x_2,x_1) \no\\ 
+ S^{+}(x_1,x_2) \Sigma^{H}(x_2,x_1) - S^{H}(x_1,x_2) \Sigma^{+}(x_2,x_1) \Big] \Big\} \no\\
+ \frac{1}{2} \text{tr}\Big[\Sigma^{>}(x_1,x_2)  S^{<}(x_2,x_1) - \Sigma^{<}(x_1,x_2)  S^{>}(x_2,x_1) \no\\ 
+ S^{<}(x_1,x_2) \Sigma^{>}(x_2,x_1) - S^{>}(x_1,x_2) \Sigma^{<}(x_2,x_1) \Big] \Big\} \,. 
\nq
We integrate the above equation over $x_1$ to find
\bq
-\Delta N_\ell &=& \int d^4x_1 d^4x_2 \Big\{ 2 \text{tr}\Big[\Sigma^{H}(x_1,x_2)  S^{+}(x_2,x_1) - \Sigma^{+}(x_1,x_2)  S^{H}(x_2,x_1) \Big] \no\\ 
&&\hspace{21mm}+ \text{tr}\Big[\Sigma^{>}(x_1,x_2)  S^{<}(x_2,x_1) - \Sigma^{<}(x_1,x_2)  S^{>}(x_2,x_1) \Big] \Big\} \,.
\label{eq:number_asymmetry}
\nq
We perform a CP transformation, where the CP properties of the lepton propagators are shown in \equaref{eq:propagator_CP} and those for the self-energy corrections preserve a similar transformation. With the help of the definition of $\Delta N_\ell$ in \equaref{eq:define_number_asymmetry}, \equaref{eq:number_asymmetry} is CP transformed to
\bq
+\Delta N_\ell &=& \int d^4x_1 d^4x_2 \Big\{ 2 \text{tr}\Big[\Sigma^{H}(x_1,x_2)  S^{+}(x_2,x_1) - \Sigma^{+}(x_1,x_2)  S^{H}(x_2,x_1) \Big] \no\\ 
&&\hspace{21mm}- \text{tr}\Big[\Sigma^{>}(x_1,x_2)  S^{<}(x_2,x_1) - \Sigma^{<}(x_1,x_2)  S^{>}(x_2,x_1) \Big] \Big\} \,.
\label{eq:number_asymmetry2}
\nq
Combining \equasref{eq:number_asymmetry}{eq:number_asymmetry2} together, we obtain 
\bq 
\Delta N_{\ell_\alpha} 
&=& - \int d^4x_1 d^4x_2 \text{tr}\Big[\Sigma^{>}_{\alpha\beta}(x_1,x_2)  S^{<}_{\beta\alpha}(x_2,x_1) - \Sigma^{<}_{\alpha\beta}(x_1,x_2)  S^{>}_{\beta\alpha}(x_2,x_1) \Big] \,,
\nq 
where the flavour indices have been included. The total lepton asymmetry is a sum of the lepton asymmetry for each single flavour, $\Delta N_\ell = \sum_{\alpha} \Delta N_{\ell_\alpha}$. 
For convenience, we will replace $\int d^4 x_1 d^4 x_2$ by $\int d^4 x d^4 r$ for our later discussion, where again $x=(x_1+x_2)/2$ and $r=x_1-x_2$. We observe that the self-energy term $\Sigma^HS^{<,>}$ and the dispersion term $\Sigma^{<,>}S^H$ do not contribute to the lepton asymmetry directly. 
We average $\Delta N_\ell$ over a volume $V $ and obtain the number density of the lepton asymmetry $\Delta n_\ell = \Delta N_\ell / V$.

%%%%%%%%%%%%%%%%%%%%%%%%%%%%%%%%%%%%%%%%%%%%%%%%%%%%%%%%%%%%%%%%%%%%%%%%%%%%%%%%%%%%%%%%%%%%%%%%%%%%%%
%%%%%%%%%%%%%%%%%%%%%%%%%%%%%%%
%%%%    CALCULATE ASYMMETRY		 %%%%%%%
%%%%%%%%%%%%%%%%%%%%%%%%%%%%%%%
\section{Calculation of the Lepton Asymmetry}\label{sec:calclep}

In \secref{sec:CTPcalc} we present  a detailed calculation of the  lepton asymmetry from the varying Weinberg operator. We follow in \secref{sec:PT} 
with a discussion of the functional form of the Weinberg operator coefficient and demonstrate that the spatial contribution to the lepton asymmetry is negligible.
We discuss thermal effects in \secref{sec:thermaleff} and finally, in \secref{sec:numerical}, we present our numerical results. 

\subsection{Lepton Asymmetry in the CTP Approach}\label{sec:CTPcalc}

\begin{figure}[t]
\centering
\includegraphics[width=0.5\textwidth]{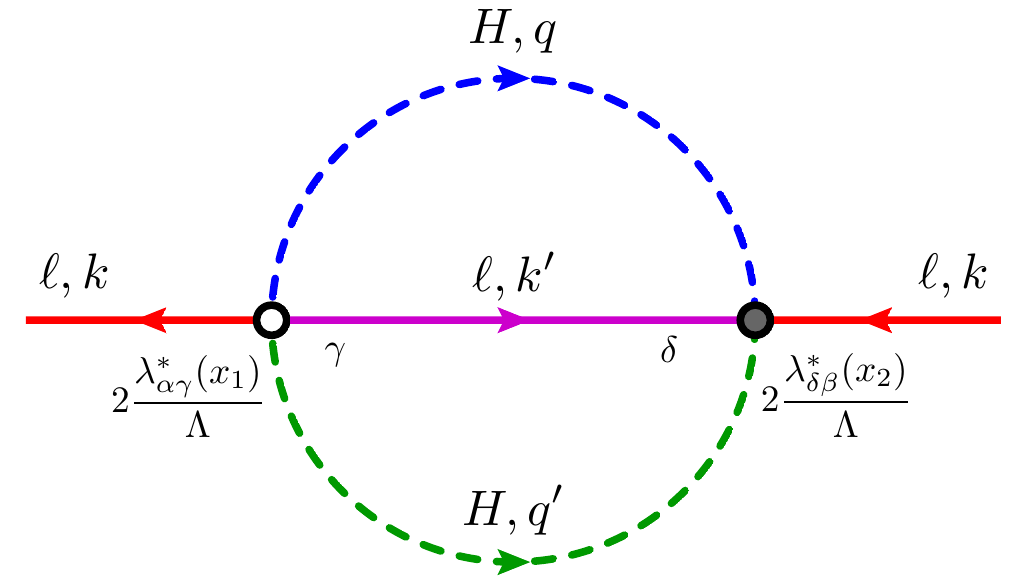}
\caption{\label{fig:Feynman}The CP-violating time-dependent two-loop contribution to the lepton self-energy induced by the Weinberg operator. }
\end{figure} 
The two-loop self-energies $\Sigma^{<,>}_{\bfk}(t_1,t_2)$ contributing to lepton asymmetry is schematically shown in  \figref{fig:Feynman}. 
The leading contribution to the lepton asymmetry enters at the two-loop level and the self-energies are given by 
\bq
\Sigma^{<,>}_{\alpha\beta}(x_1,x_2) = 3\times \frac{4}{\Lambda^2} \sum_{\gamma\delta} \lambda_{\alpha\gamma}^*(x_1) \lambda_{\delta\beta}(x_2) S^{>,<}_{\gamma\delta} (x_2, x_1) \Delta^{>,<} (x_2,x_1) \Delta^{>,<} (x_2,x_1) \,,
\nq 
where the factor $3$ comes from the $SU(2)_L$ gauge space. 
To simplify our discussion, we ignore the differing flavours  of leptons, i.e., the different thermal widths of the charged leptons. These differences arise from the different SM Yukawa couplings of $e$, $\mu$ and $\tau$ and at sufficiently high temperatures the leptonic propagators may be well approximated to be indistinguishable  and hence we apply the one-flavoured approximation, $S^{<,>}_{\alpha\beta} = S^{<,>} \delta_{\alpha\beta}$. Using this simplification, we  obtain the total lepton asymmetry summed for all 3 flavours as 
\bq
\begin{aligned}
\Delta N_\ell = - \frac{12}{\Lambda^2} \int d^4x d^4r \text{tr}[\lambda^*(x_1) \lambda(x_2)] \times
\Big\{ &\text{tr}\big[ S^{<} (x_2,x_1) S^{<}(x_2,x_1) \big] \Delta^{<} (x_2,x_1) \Delta^{<} (x_2,x_1) \\
 -& \text{tr}\big[ S^{>} (x_2,x_1) S^{>}(x_2,x_1) \big] \Delta^{>} (x_2,x_1) \Delta^{>} (x_2,x_1) \Big\} \,, 
\end{aligned}
\nq
where the trace of $\lambda$ and that of lepton propagators are understood to be  performed in the flavour space and the spinor space respectively.

We perform the following   Fourier transformation
\bq
\Delta N_\ell &=& - \frac{12}{\Lambda^2} \int d^4x d^4r\, (-i)\text{tr} [\lambda^*(x_1) \lambda(x_2)] \cM\,,
\nq
and  introduce a pure propagator function $\cM$, given by
\bq
\begin{aligned}
\cM = &i \!\int\frac{d^4k}{(2\pi)^4} \frac{d^4k'}{(2\pi)^4} \frac{d^4q}{(2\pi)^4} \frac{d^4q'}{(2\pi)^4}  e^{iK\cdot (-r)} \\
& \times \left\{ \text{tr} [ S^{<}_{k}(x) S^{<}_{k'}(x) ] \Delta^{<}_{q}(x) \Delta^{<}_{q'}(x) \! - \! \text{tr} [ S^{>}_{k}(x) S^{>}_{k'}(x) ] \Delta^{>}_{q}(x) \Delta^{>}_{q'}(x) \right\}\!,
\end{aligned}
\nq
where $K=k+k'+q+q'$. 
As the temperature of the PT is much higher than the EW scale, it is a sufficiently good approximation to 
 assume thermal distributions of the propagators on the RHS of the above equation (for the non-equibrium contribution, see the discussion in \appref{sec:EWBG}). The space-independent propagators $\Delta^{<}_{q}$, $\Delta^{<}_{q'}$, $S^{<}_{k}$ and $S^{<}_{k'}$ in \equaref{eq:propagator_equilibrium} can be directly taken into the above equation. Then, the propagator combination $\Delta_q^<\Delta_{q'}^<S_k^<S^<_{k'} - \Delta_q^>\Delta_{q'}^>S_k^>S^>_{k'}$ is proportional to
\bq
&& 
[ \vartheta(- k^0) - f_{F, |k^0|} ][ \vartheta(- k'^0) - f_{F, |k^{\prime0}|} ] 
[ \vartheta(- q^0) + f_{B, |q^0|} ][ \vartheta(- q'^0) + f_{B, |q^{\prime0}|} ]\no\\
&-& 
[ \vartheta(+ k^0) - f_{F, |k^0|} ][ \vartheta(+ k'^0) - f_{F, |k^{\prime0}|} ] 
[ \vartheta(+ q^0) + f_{B, |q^0|} ][ \vartheta(+ q'^0) + f_{B, |q^{\prime0}|} ]\,,
\nq 
which is obviously an odd function under the transformation $\{q,q',k, k'\} \leftrightarrow -\{q,q',k, k'\}$. 
With the help of this property, it is straightforward to obtain 
\bq\label{eq:pprop}
\cM &=& \frac{1}{2}\left\{ \cM + \cM|_{\{q,q',k, k'\} \to -\{q,q',k, k'\}} \right\} \no\\
&=&\int\frac{d^4k}{(2\pi)^4} \frac{d^4k'}{(2\pi)^4} \frac{d^4q}{(2\pi)^4} \frac{d^4q'}{(2\pi)^4} \text{Im}\left\{ e^{iK\cdot r}\right\} \Big[ \text{tr} [ S^{<}_{k} S^{<}_{k'} ] \Delta^{<}_{q} \Delta^{<}_{q'} - \text{tr} [ S^{>}_{k} S^{>}_{k'} ] \Delta^{>}_{q} \Delta^{>}_{q'} \Big],
\label{eq:cM}
\nq 
where we note that $\mathcal{M}$ is odd under the exchange $x_1\leftrightarrow x_2$. Eventually, we simplify the lepton asymmetry to 
\bq
\Delta N_\ell &=& - \frac{12}{\Lambda^2} \int d^4x d^4r\text{Im}\left\{\text{tr} \left[\lambda^*(x_1) \lambda(x_2)\right]\right\} \cM  \,,
\nq 
where $x = (x_1+x_2)/2$ and $r=x_1-x_2$ represent the average and relative of coordinates $x_1$ and $x_2$, respectively. 
The lepton asymmetry has been factorised into two parts: $\cM$ is a function the propagators and $\text{Im}\left\{\text{tr} \left[\lambda^*(x_1) \lambda(x_2)\right]\right\}$ contains the couplings.

As previously mentioned,  we assume temperatures much higher than the EW scale, and therefore all propagators for the Higgs and leptons in $\cM$ are in thermal equilibrium. Thus KMS relations for Wightman propagators $\Delta^{>}_{q} = e^{\beta q^0} \Delta^{<}_{q}$, $S^{>}_{k} = - e^{\beta k^0} S^{<}_{k}$ are satisfied. 
We would like to stress although the KMS relation is satisfied, the propagator function  $\text{tr} [ S^{<}_{k} S^{<}_{k'} ] \Delta^{<}_{q} \Delta^{<}_{q'} - \text{tr} [ S^{>}_{k} S^{>}_{k'} ] \Delta^{>}_{q} \Delta^{>}_{q'}$ of \equaref{eq:pprop} does not vanish as the momenta of the four propagators does not equal zero as shall see shortly. 
Using the 
 tree-level propagator  given in \equasref{eq:Higgs_propagator}{eq:lepton_propagator} with distribution functions in \equaref{eq:equilibrium} and  assuming thermal equilibrium in the rest frame of the plasma, we can prove $\cM$ is an even function of $\bfr$. To do so we perform the following parity transformation for $\cM$: 
\bq
r \to r^P = (r^0,-\bfr)\,, \quad k_n \to k_n^P=(k^0_n, -\bfk_n)\,,
\label{eq:parity}
\nq 
where $k_n$ represents each of $k,k'q,q'$. 
Note that the tree-level $\Delta_q^{<,>}$ is invariant under the spatial parity transformation, $\Delta_q^{<,>}=\Delta_{q^P}^{<,>}$. Although $S^{<}_{k}$ is not invariant under $k \to k^P$, the trace is: $ \text{tr} [ S^{<,>}_{k^P} S^{<,>}_{k^{\prime P}} ] =  \text{tr} [ S^{<,>}_{k} S^{<,>}_{k'} ]$. From these properties, we directly prove that $\cM$ is invariant under the parity transformation as shown in \equaref{eq:parity} and therefore $\cM$ is an even function of $\bfr$. Including the SM loop corrections, we will obtain thermal damping effect and dispersion relations which will be discussed in the next section. The SM loop corrections modify the tree-level propagators but do not change the properties of $\cM$ which  is an even function of $\bfr$ because no spatial-specific interactions have been included in the SM. This schematic discussion demonstrates that although the Weinberg operator is
\textbf{spacetime-dependent} only the temporal component contributes to the final lepton asymmetry.  This will be further elucidated in \secref{sec:PT}.

In summary,  to generate a lepton asymmetry it is necessary to include a CP-violating spacetime-varying Weinberg operator. If the coupling is \textbf{spacetime-independent}, we immediately arrive at the 4-momentum conservation $K\equiv q+q'+k+k'=0$ from the integration $\int d^4r$, and obtain $\Delta^{<}_{q} \Delta^{<}_{q'} \text{tr} [ S^{<}_{k} S^{<}_{k'} ] - \Delta^{>}_{q} \Delta^{>}_{q'} \text{tr} [ S^{>}_{k} S^{>}_{k'} ] = 0$, and thus  $\Delta N_\ell$ vanishes with the help of the KMS relation. 
To generate a non-zero $\Delta N_\ell$, CP violation in the varying Weinberg operator is also a necessary condition. This comes from the imaginary part of $\text{tr} [\lambda^*(x_1) \lambda(x_2)]$ and leads to the CP violation for the lepton/anti-lepton production and annihilation processes. 

%%%%%%%%%%%%%%%%%%%%%%%%%%%%%%%
%%%%    			PT			  		%%%%%%%
%%%%%%%%%%%%%%%%%%%%%%%%%%%%%%%

\subsection{Simplification of the Phase Transition Contribution}\label{sec:PT}
In general, the dynamics of a PT are complicated. 
To simplify our discussion, we will only consider the simplest case that only a single scalar $\phi\equiv \phi_1$ is involved in the phase transition and the coefficient of the Weinberg operator is linearly dependent upon $\phi$ as $\lambda_{\alpha\beta} = \lambda^0_{\alpha\beta} + \lambda^1_{\alpha\beta} \phi / v_{\phi}$. We note that we provide an extensive discussion of the multi scalar PT  in Appendix \ref{sec:PT2}. 
As we have assumed a first-order PT throughout this work, $\lambda(x)$ is determined by the property of the bubble wall. 
We treat the scalar field $\phi$ as a thermal bath with temperature $T=1/\beta$. 
The system begins its evolution at $t=-\infty$ in Phase I, $\langle \phi \rangle=0$. 
After a certain period, $\langle \phi \rangle$ varies from 0 to $v_\phi$ and the system enters Phase II. During the phase transition, the spacetime-dependent scalar EEV $\langle \phi(x) \rangle$ can be parametrised as $\langle \phi(x) \rangle = f_1(x') v_\phi$, where $f(x')$ represents the EEV shape smoothly varying from $0$ to $1$ for $x'\equiv x^3 + v_w x^0$ running from $-\infty$ to $+\infty$. Typical examples of the bubble profiles are given in Appendix \ref{sec:EEVprofle}. 
 As a consequence, the coupling $\lambda(x)$ is given by 
\bq
\lambda(x)=\lambda^0+\lambda^1 f_1(x')\,. \label{eq:EEVprofile}
\nq 
Typical examples of the bubble profiles are given in Appendix \ref{sec:EEVprofle}. 
Then, $\text{Im}\left\{ \text{tr} \left[\lambda^*(x_1) \lambda(x_2)\right] \right\}$ is simplified to 
\bq
\text{Im}\left\{ \text{tr} \left[\lambda^*(x_1) \lambda(x_2)\right] \right\} = \text{Im}\left\{ \text{tr} \left[\lambda^0 \lambda^{1*}\right] \right\} [f_1(x_1') - f_1(x_2')]\,.
\nq
By assuming a small difference $r' \equiv x'_1-x'_2$, the integration
\bq \label{eq:int_xp}
\int_{-\infty}^{+\infty} d x' [ f_1(x'+ r'/2) - f_1(x'- r'/2) ] \approx \int_{-\infty}^{+\infty} d x' \partial_{x'}f_1(x') r' = r',
\nq
is independent of the scalar EEV profile in the wall, and certainly independent of the wall thickness $L_w$\footnote{Taking the examples in Appendix \ref{sec:EEVprofle}, one can check its validity. However, this result is independent from these special profiles. }. Making use of the above integration, we arrive at 
\bq
\int d^4x \text{Im}\{\text{tr}[\lambda^*(x_1)\lambda(x_2)] \} &=& \text{Im}\{ \text{tr} [ \lambda^{0} \lambda^* ] \} \Big(r^0+\frac{r^3}{v_w}\Big) V \,,
\label{eq:int_x}
\nq
where $\int d^3\bfx = V$ and $\text{Im}\{ \text{tr} [ \lambda^{0} \lambda^{1*} ] \} = \text{Im}\{ \text{tr} [ \lambda^{0} \lambda^* ] \}$ have been used. In the single scalar case, the exact functional form of the scalar EEV profile is not important.  

From \equaref{eq:int_x}, we see that the number density of the lepton asymmetry becomes $\Delta n_\ell = \Delta n_\ell^\text{I} + \Delta n_\ell^\text{II}$, with
\bq 
\Delta n_\ell^\text{I} &=& - \frac{12}{\Lambda^2} \text{Im}\{ \text{tr} [ \lambda^{0} \lambda^* ] \} \int d^4r\, r^0 \,\mathcal{M} \,, \no\\
\Delta n_\ell^\text{II} &=& \frac{12}{v_w\Lambda^2} \text{Im}\{ \text{tr} [ \lambda^{0} \lambda^* ] \} \int d^4r\, r^3\, \mathcal{M}\,,
\label{eq:LeptonAsymm}
\nq
where $\Delta n_\ell^\text{I}$ and $\Delta n_\ell^\text{II}$ represent the time-dependent and space-dependent lepton asymmetry \textbf{in the rest plasma frame} respectively. They correspond to integrations along $r^0$ and $r^3/v_w$, respectively. 
We comment that the space-dependent integration $\Delta n_\ell^\text{II}$ vanishes due to our assumption of thermal equilibrium of the Higgs and leptons as shown in \equaref{eq:LeptonAsymm}. This is because in thermal equilibrium, there are no preferred momentum and space directions for the propagators. We perform the following parity transformation: 
\bq
r \to r^P = (r^0,-\bfr)\,, \quad k_n \to k_n^P=(k^0_n, -\bfk_n)\,,
\label{eq:parity}
\nq 
where $k_n$ represents each of $k,k'q,q'$. 
Note that $\Delta_q^{<,>}$ is invariant under the spatial parity transformation, $\Delta_q^{<,>}=\Delta_{q^P}^{<,>}$. Although $S^{<}_{k}$ is not invariant under $k \to k^P$, but $ \text{tr} [ S^{<,>}_{k^P} S^{<,>}_{k^{\prime P}} ] =  \text{tr} [ S^{<,>}_{k} S^{<,>}_{k'} ]$ is satisfied. Therefore $\cM$ is invariant under the parity transformation in Eq. \eqref{eq:parity}. In other words, $\cM$ is an even function of $\bfr$ and consequently the space-dependent integration $\int d^4r \,r^3 \mathcal{M}$ vanishes. 
The propagators are not invariant under the time parity transformation $r\to -r^P$ and $k_n \to -k_n^P$ due to the statistical factor. Thus, $\mathcal{M}$ is not an even function of $r^0$, and the time-dependent integration $\int d^4 r \, r^0 \mathcal{M}$ does not vanish. 
Thus, the final lepton asymmetry in the single scalar case is only time-dependent, $\Delta n_\ell = \Delta n_\ell^{\text{I}}$, i.e.,
\bq
\Delta n_\ell = - \frac{12}{v_H^4} \text{Im}\{ \text{tr} [ m_\nu^{0} m_\nu^* ] \} \int d^4r\, y \,\mathcal{M}\,,
\label{eq:lepton_asymmetry_f1}
\nq
where $r^0$ is re-written as $y$ for convenience. 

Based on the result in \equaref{eq:lepton_asymmetry_f1}, we conclude that the lepton asymmetry in the single scalar case is determined by two parts: 1) the neutrino mass combination $\text{Im}\{ \text{tr} [ m_\nu^{0} m_\nu^* ] \}$ and 2) the time-dependent loop integration $\int d^4r\, y \,\mathcal{M}$. Bearing in mind \equaref{eq:EEVprofile}, the dependence upon $\text{Im}\{ \text{tr} [ m_\nu^{0} m_\nu^* ] \}$ means that the lepton asymmetry depends only on an initial non-zero value of the coefficient of the Weinberg operator with coefficient $\lambda^0$ and a relative phase between $\lambda^0$ and the final value $\lambda$. In other words, it does not depend upon the profile of the $\phi$ EEV,  $f(x')$ in \equaref{eq:EEVprofile}, i.e., the property of the bubble wall, no matter the thin wall or thick wall. However, this conclusion does not fully hold when extending to the multiple scalar case. We leave the relevant discussion to the next section. As will be shown there, the lepton asymmetry is non-trivially determined by the properties of the bubble wall. 
The second interesting point is we have proved that in the rest plasma frame, only the time-dependent loop integration $\int d^4r\, y \,\mathcal{M}$ is involved in leptogenesis. We will prove in the next section that this conclusion is true in the more general multiple scalar case.

%%%%%%%%%%%%%%%%%%%%%%%%%%%%%%%
%%%%   THERMAL EFFECTS  			%%%%%%%
%%%%%%%%%%%%%%%%%%%%%%%%%%%%%%%
\subsection{Inclusion of Thermal Effects}\label{sec:thermaleff}

In the previous section we encountered the time-dependent propagator integration $\int d^4r y \cM$, where the Higgs and lepton propagators are assumed to be in thermal distribution in $\cM$. Although, the tree-level propagators have been given in \equasref{eq:Higgs_propagator}{eq:lepton_propagator}, they are not enough to guarantee a convergence result for the integration. This integration is strongly dependent upon the thermal properties of the particles, specifically dependent upon the damping rate. 

Taking the loop correction into account, the resumed Wightman propagators of leptons and the Higgs in thermal distribution can be expressed in the Breit-Wigner form \cite{Aurenche:1991hi,vanEijck:1992mq,Millington:2012pf}: 
\bq
\Delta^{<,>}_{q}&=& \frac{-2\varepsilon(q^0) \Imag \Pi^R_q}{[q^2-\Real \Pi^{R}_q]^2 + [\Imag \Pi^R_q]^2} \Big\{ \vartheta(\mp q^0) + f_{B, |q^0|}(x) \Big\} \,,\no\\
S^{<,>}_{k} &=& \frac{-2\varepsilon(k^0) \Imag \Sigma^{R\,2}_k}{[k^2-\Real \Sigma^{R}_q]^2 + [\Imag \Sigma^{R\,2}_q]^2} \Big\{ \vartheta(\mp k^0) - f_{F, |q^0|}(x) \Big\} P_L k\!\!\!/ P_R \,,
\label{eq:thermal_propagator}
\nq
where $\varepsilon(q^0)=\vartheta(q^0)-\vartheta(-q^0)$, $\Pi^R_{q}$, $\Sigma^R_k$ are retarded self-energies of the Higgs and leptons respectively. Replacing the tree-level propagators with \equaref{eq:thermal_propagator}, we recover $\cM$ in \equaref{eq:cM}. All equilibrium propagators are spacetime-independent. In the limit $\Pi^R_q, \Sigma^R_k \to 0$ and by using the representation of the delta function
\bq
\delta(a) = \frac{1}{\pi} \lim_{\gamma\to 0} \frac{\gamma}{a^2+\gamma^2}\,,
\nq
we recover the free propagators in \equaref{eq:propagator_equilibrium} with equilibrium distributions. 
The thermal masses and widths are defined from the real and imaginary parts of self energies as $\Real \Pi = m_{\text{th}}^2$ and $\Imag \Pi = 2 m_{\text{th}} \gamma$ respectively and therefore  \equaref{eq:thermal_propagator} becomes
\bq
\Delta^{<,>}_{q} &\approx& \Big(\coth\frac{\beta q^0}{2} \mp 1 \Big)\frac{2 q^0 \gamma_H}{[(q^0)^2- |\bfq|^2 -m_{H,\text{th}}^2 ]^2 + (2q^0 \gamma_H)^2} \,, \no\\
S^{<,>}_{k} &\approx& \Big(\tanh\frac{\beta k^0}{2} \mp 1 \Big)\frac{2 k^0 \gamma_\ell}{[(k^0)^2- |\bfk|^2 -m_{\ell,\text{th}}^2 ]^2 + (2k^0 \gamma_\ell)^2} P_L k\!\!\!/P_R \,. 
\label{eq:thermal_propagator_v2}
\nq
As discussed earlier, we do not distinguish thermal corrections to different flavours. All lepton doublets have the same thermal widths,
$\gamma_e=\gamma_\mu=\gamma_\tau\equiv\gamma_{\ell, \bfk}$, which is a function of the momentum $\bfk$. In the SM, the processes which dominantly contribute to the leptonic thermal widths are EW gauge interactions and the thermal width at zero momentum $\gamma_{\ell,\bfk=0} \approx 6/(8\pi)g^2 T \approx 0.1T$ \cite{bellac}, where $g$ is the $SU(2)_L$ gauge coupling. For the Higgs, both EW gauge interaction and the top quark Yukawa coupling contribute to the Higgs thermal width, 
thus $\gamma_{H,\bfq=0} \approx 3/32\pi g^2 T + 3/8\pi y^2_t T\approx 0.1 T$ \cite{Biro:1996gf} where $y_t$ is the top quark Yukawa coupling. In this paper, we shall fix $\gamma_{\ell}$ and $\gamma_H$ at certain constant values. For non-vanishing momentum, the thermal width is in general momentum-dependent and  BSM interactions may modify their values. These effects may quantitatively modify the final generated lepton asymmetry and will be discussed elsewhere. 

In the following, we will calculate $d^4r y \cM$ using linear response limit. Such a treatment originates from the time-dependent coupling of the Weinberg operator.
The latter corresponds to energy transfer between particles and the background which leads to energy non-conservation of particles \cite{Millington:2012pf}. In order to deal with this scenario, we simplify our discussion in the narrow-width limit. 
The final result has already been shown in our former work Ref.~\cite{Pascoli:2016gkf}.

Firstly, we would like to integrate over the time-difference $y\equiv r^0$. This can be done with the help of the following Fourier transformations 
\bq
\Delta^{<,>}_{\bfq} (t_1,t_2) &=& \int \frac{d q^0}{2\pi} e^{-i q^0 y} \Delta^{<,>}_q \,, \no\\
S^{<,>}_{\bfk} (t_1,t_2) &=& \int \frac{d k^0}{2\pi} e^{-i k^0 y} S^{<,>}_k \,. 
\nq 
Since the width $\gamma_{H,\bfq}, \gamma_{H,\bfk}\ll T$ we may safely ignore the terms  $\mathcal{O}(\gamma_{H,\bfq}^2/T^2,\gamma_{\ell,\bfk}^2/T^2)$ and  we find the propagators for the Higgs and leptons as $\Delta^{<,>}_{\bfq} (t_1,t_2) = \Delta^{T}_{\bfq} (t_1,t_2)+ \Delta^{0<,>}_{\bfq} (t_1,t_2)$ and $S^{<,>}_{\bfk} (t_1,t_2) = S^{T}_{\bfk} (t_1,t_2) + S^{0<,>}_{\bfk} (t_1,t_2)$, where
\bq
\Delta^{T}_{\bfq} (t_1,t_2) &=& f_{B,|\bfq|}\frac{1}{2\omega_\bfq}(e^{i\omega_\bfq y}+ e^{-i\omega_\bfq y}) e^{-\gamma_{\ell,\bfk} |y|} \no\\ 
\Delta^{0<,>}_{\bfq} (t_1,t_2) &=& \frac{1}{2\omega_\bfq} e^{\pm i\omega_\bfq y -\gamma_{\ell,\bfk} |y|} \,, \no\\
S^{T}_{\bfk} (t_1,t_2) &=& f_{F,|\bfk|} \frac{1}{2} (P_L \hat{k}\!\!\!/_+ e^{i\omega_\bfk y} + P_L \hat{k}\!\!\!/_- e^{-i\omega_\bfk y}) e^{-\gamma_{\ell,\bfk} |y|} \no\\
S^{0<,>}_{\bfk} (t_1,t_2) &=& - \frac{1}{2} P_L \hat{k}\!\!\!/_\pm e^{\pm i\omega_\bfk y-\gamma_{\ell,\bfk} |y|},
\nq 
and $\omega_\bfq = \sqrt{m_{H,\text{th}}^2+\bfq^2}$, $\omega_\bfk = \sqrt{m_{\ell,\text{th}}^2+\bfk^2}$, $\hat{k}\!\!\!/_\pm = \pm\gamma_0 + \hat{\bfk}\cdot \vec{\gamma}$ with $\hat{\bfk}\equiv \bfk/\omega_\bfk$ \cite{Anisimov:2010dk}. As expected, the thermal components, labeled by $T$, are the same for $<$ and $>$, and the zero temperature parts, labeled by $0$, are different. 
After performing these Fourier transformations and integrating over the spatial component, $\int d^3\bfr$, we obtain a Delta function $\delta^{(3)}(\bfk+\bfk'+\bfq+\bfq')$. This corresponds to the three-dimensional momentum conservation~\footnote{Note that the spatial integration $\int d^4 r r^3 \mathcal{M}$ can lead to momentum non-conservation along $r^3$ direction. This effect, as discussed above, does not contribute to the lepton asymmetry.}. We integrate over $\bfk'$ and simplify the time-integration to 
\bq 
\int d^4r\, y \cM
=  2 \int  \frac{d^3\bfk}{(2\pi)^3}\frac{d^3\bfq}{(2\pi)^3} \frac{d^3\bfq'}{(2\pi)^3} \int_{-\infty}^{+\infty} dy\, y M \,,
\label{eq:LeptonAsymm_I}
\nq
where 
\bq 
\begin{aligned}\label{eq:traceexp}
M&= \text{tr} [ S^{<}_{\bfk}(t_1,t_2) S^{<}_{\bfk'}(t_1,t_2) ] \Delta^{<}_{\bfq}(t_1,t_2) \Delta^{<}_{\bfq'}(t_1,t_2)\\
& -  \text{tr}[S^{>}_{\bfk}(t_1,t_2) S^{>}_{\bfk'}(t_1,t_2) ] \Delta^{>}_{\bfq}(t_1,t_2) \Delta^{>}_{\bfq'}(t_1,t_2),
\end{aligned}
\nq
and $\bfk'$ is fixed at $\bfk'=-(\bfk+\bfq+\bfq')$. \\

Following \appref{sec:ME}, we represent the propagators as
\bq \label{eq:eqprop}
\Delta^{<,>}_{\bfq} (t_1,t_2) &=& \frac{\cos(\omega_\bfq y^\mp)}{2 \omega_\bfq \sinh(\omega_\bfq\beta/2)} e^{-\gamma_{H,\bfq} |y|} \,, \no\\
S^{<,>}_{\bfk} (t_1,t_2) &=& - P_L \frac{\gamma^0 \cos(\omega_\bfk y^\mp) + i \vec{\gamma}\cdot \hat{\bfk}\sin(\omega_\bfk y^\mp)}{2 \cosh(\omega_\bfk\beta/2)} e^{-\gamma_{\ell,\bfk} |y|}\,,
\nq 
where $y^\mp \equiv y\mp i \beta/2$. Then, $M$ is simplified to
\bq
\label{eq:ME2}
M&=&\frac{\text{Im} \{ [c(\omega_\bfk y^-)c(\omega_{\bfk'}y^-)+\hat{\bfk}\cdot \hat{\bfk}' s(\omega_{\bfk} y^-) s(\omega_{\bfk'} y^-)] c(\omega_\bfq y^-)c(\omega_{\bfq'}y^-) \}}
{8 \omega_{\bfq} \omega_{\bfq'} ch(\omega_{\bfk} \beta/2) ch(\omega_{\bfk'} \beta/2) sh(\omega_{\bfq} \beta/2) sh(\omega_{\bfq'} \beta/2)}  e^{-\gm|y|} \,, 
\label{eq:M}
\nq
where $\gm=\gamma_{H,\bfq}+\gamma_{H,\bfq'}+\gamma_{\ell,\bfk}+\gamma_{\ell,\bfk'}$ 
and we have changed to the notation $\cos\equiv c, \sin\equiv s$, $\cosh\equiv ch$ and $\sinh\equiv sh$ for brevity.
Note that some additional details may be found in \appref{sec:ME}.
In this form, we can straightforwardly prove that $M$ is an odd function of $y$, and we can integrate over $y$ in the following way: 
\bq 
\begin{aligned}
\hspace*{-5mm}&\int_{-\infty}^{+\infty} dy y M = 2 \int_{0}^{+\infty} dy y M  \\
\hspace*{-5mm}=&  2 \int_{0}^{+\infty} dy y \frac{\text{Im} \{ [c(\omega_{\bfk}y^-)c(\omega_{\bfk'}y^-)+\hat{\bfk}\cdot \hat{\bfk}' s(\omega_{\bfk} y^-) s(\omega_{\bfk'} y^-)] c(\omega_{\bfq}y^-)c(\omega_{\bfq'}y^-) \}}
{8\omega_{\bfq} \omega_{\bfq'} ch(\omega_{\bfk} \beta/2) ch(\omega_{\bfk'} \beta/2) sh(\omega_{\bfq} \beta/2) sh(\omega_{\bfq'} \beta/2) } e^{-\gm y}  \\
\hspace*{-5mm}=& \sum_{\eta_2,\eta_3,\eta_4=\pm1} \frac{ [1- \eta_2\hat{\bfk}\cdot \hat{\bfk}'] K_{\eta_2\eta_3\eta_4} \gm}{16 \omega_{\bfq} \omega_{\bfq'} (K_{\eta_2\eta_3\eta_4}^2+\gm^2)^2}
\frac{ sh(\beta K_{\eta_2\eta_3\eta_4} /2)}{ ch(\omega_{\bfk} \beta/2) ch(\omega_{\bfk'} \beta/2) sh(\omega_{\bfq} \beta/2) sh(\omega_{\bfq'} \beta/2)},
\label{eq:inteM}
\end{aligned}
\nq
where $K_{\eta_2\eta_3\eta_4} = \omega_{\bfk}+\eta_2 \omega_{\bfk'}+\eta_3 \omega_{\bfq}+\eta_4 \omega_{\bfq'}$. 

In the semi-classical point of view, each $K_{\eta_2\eta_3\eta_4}$ corresponds to the energy transfer from the bubble wall to different processes by the Weinberg operator, in detail,
\bq
K_{+++}: & \quad \text{ vacuum energy transfer to } \quad & 0 \to \ell \ell H H \no\\
K_{++-} ~\&~ K_{+-+}: & \cdots & H^* \to \ell \ell H \no\\
K_{+--}: &\cdots & H^* H^* \to \ell \ell \no\\
K_{-++}: &  \cdots & \overline{\ell} \to \ell H H \no\\
K_{-+-} ~\&~ K_{--+}: & \cdots & \overline{\ell} H^* \to \ell H \no\\
K_{---}: & \cdots & \overline{\ell} H^* H^* \to \ell.
\nq
During the PT, the false vacuum, which carries higher energy than the true vacuum, releases energy to the true vacuum. This energy is partially transferred to the kinetic energy of the lepton and Higgs via the Weinberg operator.
In the limit of zero energy transfer, $K_{\eta_2\eta_3\eta_4}\to 0$, the integration in Eq.~\eqref{eq:inteM} is zero and no lepton asymmetry is generated. 
This is to be anticipated as the distribution functions of the leptons and Higgs remains thermal.
This transfer of energy between the leptons, Higgs and bubble wall can be understood in terms of the interactions
between these particles with the scalar field, $\phi$. 
Deep inside the bubble the scalar is massive, while in the symmetric phase the scalar remains massless and rather  obviously the scalar mass varies across the bubble wall. For a very fast moving bubble wall expansion, these scalars in the bubble wall are highly off-shell because of the large spacetime  gradient of the VEV in the bubble wall. The momentum of the off-shell scalars may be transferred
to the leptonic doublets and Higgses via scatterings mediated by the dynamically-realised Weinberg operator. Here, we do not fix the energy transfer but assume an upper bound of the energy transfer around the temperature. We address this issue in details in Appendix~\ref{sec:oscillating}. 

These scatterings may cause the necessary perturbations of the leptons, anti-leptons and Higgs distribution functions from equilibrium.
There will be interference between this process and those mediated by the dimension-five operator which will result in a non-zero lepton asymmetry. %

The energy transfer is, in principle, not free, but dependent upon interactions between the scalar and other particles. In this paper, we only include the effective interaction in the Weinberg operator, which is very weak. If any additional interactions of the scalar with lepton or with the Higgs are stronger than it, the energy transfer will be determined by the new interactions. Furthermore, the lepton and Higgs released from the bubble wall may be off-shell and followed up with transition radiation \cite{Bodeker:2017cim}, which complicates the energy transfer from the bubble wall to the plasma. Instead of discussing these processes in details, we simplify their contributions by adding an upper bound $K_{\text{cut}}$, i.e., a cut of the transfer energy $|K_{\eta_2\eta_3\eta_4}|\lesssim K_\text{cut}$ for all $K_{\eta_2\eta_3\eta_4}$. This is realised by including  $\varepsilon(K_{\eta_2\eta_3\eta_4}, K_{\text{cut}}) \equiv \vartheta(K_{\eta_2\eta_3\eta_4}+K_{\text{cut}}) - \vartheta(K_{\eta_2\eta_3\eta_4}-K_{\text{cut}})$ in \equaref{eq:inteM}, where $\vartheta(x)$ is the Heaviside function.  We \emph{estimate} the maximum of this momentum transfer to be of  the order of the temperature \emph{i.e.} $K_{\text{cut}}\sim \mathcal{O}\left( 1\right)T$ and relegate a more detailed calculation for future work. This simplified treatment is also supported by the numerical calculation: as 
we vary $K_{\text{cut}}$ around $T$ and observe that the integration, shown later in \equaref{eq:Fxgammaxcut}, is not strongly dependent upon the exact value of $K_{\text{cut}}$. However, for $K_{\rm cut} \gg T$, the phase space is  enlarged, and our calculation is not applied. 

To calculate the momentum integration, we follow the technique in \cite{Anisimov:2010dk}. Assuming $\bfp\equiv \bfk-\bfq$, we replace the momentum integration $d^3\bfq d^3 \bfq'$ to $d^3\bfk' d^3\bfp$, where $\bfp= \bfk'+\bfq'$ holds obviously and 
\bq
\hat{\bfk}\cdot \hat{\bfk}' &=& \frac{(|\bfk|^2+|\bfp|^2-|\bfq|^2)(|\bfk'|^2+|\bfp|^2-|\bfq'|^2)}{4 \omega_{\bfk} \omega_{\bfk'} |\bfp|^2}  \,.
\nq
With the help of the following parametrisation
\bq
\bfp &=& |\bfp|(0,0,1)\,,\no\\
\bfk &=& |\bfk|(\sin\theta,0,\cos\theta)\,,\no\\
\bfk' &=& |\bfk'|(\sin\theta'\cos\varphi',\sin\theta'\sin\varphi',\cos\theta')\,,
\nq 
we derive 
\bq
\hspace*{-5mm}&&\int \frac{d^3\bfk}{(2\pi)^3} \frac{d^3\bfq}{(2\pi)^3} \frac{d^3\bfq'}{(2\pi)^3} = \int \frac{d^3\bfk}{(2\pi)^3} \frac{d^3\bfk'}{(2\pi)^3} \frac{d^3\bfp}{(2\pi)^3} \,, \no\\
\hspace*{-5mm}&&= \frac{2}{(2\pi)^6} \int_0^{+\infty} d|\bfp| \int_0^{+\infty} |\bfk| d|\bfk| \int_0^{+\infty} |\bfk'| d|\bfk'| \int_{\big||\bfk|-|\bfp|\big|}^{|\bfk|+|\bfp|} |\bfq| d|\bfq| \int_{\big||\bfk'|-|\bfp|\big|}^{|\bfk'|+|\bfp|} |\bfq'| d|\bfq'| \,,
\label{eq:}
\nq
where 
\bq
|\bfq|^2=|\bfk|^2+|\bfp|^2-2|\bfk||\bfp|\cos\theta \,, \quad
|\bfq'|^2=|\bfk'|^2+|\bfp|^2-2|\bfk'||\bfp|\cos\theta'
\nq 
have been used. 

Analytically, the integration can be represented by a five-variable integration. 
To simplify our discussion, we neglect the contribution from the small thermal mass, which is of the order $gT$ for the gauge coupling $g$ or $y_t T$ for the top quark Yukawa coupling, i.e., setting $\omega_{\bfk} = |\bfk|$, $\omega_{\bfq} = |\bfq|$. As mentioned previously, we neglect the momentum-dependent contribution of the thermal width, i.e., $\gm = 2(\gamma_H+\gamma_\ell)$ is taken to be constant.  We rescale the momentum in the unit of temperature, $x_1=|\bfk|\beta/2$, $x_2=|\bfk'|\beta/2$, $x_3=|\bfq|\beta/2$, $x_4=|\bfq'|\beta/2$, $x_\gamma = \gm\beta/2$, $x=|\bfp|\beta/2$, and $X_{\eta_2\eta_3\eta_4}=x_1+\eta_2x_2 + \eta_3 x_3 + \eta_4 x_4$. Finally, we arrive at  the expression
\bq
\int d^4r\, y \mathcal{M} = \frac{4\,T^5}{(2\pi)^6} F(x_\gamma, x_\omega),
\nq
where
\bq \label{eq:Fxgammaxcut}
\hspace*{-6mm}F(x_\gamma, x_\text{cut}) = 
\int_0^{+\infty} dx \int_0^{+\infty} x_1 dx_1 \int_0^{+\infty} x_2 dx_2 \int_{|x_1-x|}^{x_1+x} dx_3 \int_{|x_2-x|}^{x_2+x} dx_4 \sum_{\eta_2,\eta_3,\eta_4=\pm1}  \no\\
\hspace*{-6mm}\times \left[1- \frac{(x_1^2+x^2-x_3^2)(x_2^2+x^2-x_4^2)}{4 \eta_2 x_1 x_2 x^2} \right] \frac{\varepsilon(X_{\eta_2\eta_3\eta_4}, x_{\text{cut}}) X_{\eta_2\eta_3\eta_4} x_\gamma \, \sinh X_{\eta_2\eta_3\eta_4} }{ (X_{\eta_2\eta_3\eta_4}^2+x_\gamma^2)^2 \cosh x_1 \cosh x_2 \sinh x_3 \sinh x_4} \,.
\nq

In the thin wall case, we directly take the propagator integration to \equaref{eq:LeptonAsymm} and obtain the lepton asymmetry as 
\bq
\Delta n_\ell &=& -
\frac{3\, \text{Im}\{\text{tr}[m_\nu^{0} m_\nu^*] \} T^5}{4 \pi^6 v_H^4 } F(x_\gamma, x_\text{cut}) \,.
\nq
We also present the lepton asymmetry distribution per momentum $\bfk$: 
\bq
L_{\bfk} &=& - \frac{3\, \text{Im}\{\text{tr}[m_\nu^{0} m_\nu^*] \} T^2}{(2\pi)^4 v_H^4 } F(x_1,x_\gamma, x_\text{cut}),
\label{eq:Lbfk}
\nq
\bq
&&F(x_1,x_\gamma, x_\text{cut}) = \frac{1}{x_1} \int_0^{+\infty} dx \int_0^{+\infty} x_2 dx_2 \int_{|x_1-x|}^{x_1+x} dx_3 \int_{|x_2-x|}^{x_2+x} dx_4 \sum_{\eta_2,\eta_3,\eta_4=\pm1}  \no\\
&&\times \left[1- \frac{(x_1^2+x^2-x_3^2)(x_2^2+x^2-x_4^2)}{4 \eta_2 x_1 x_2 x^2} \right] \frac{\varepsilon(X_{\eta_2\eta_3\eta_4}, x_{\text{cut}}) X_{\eta_2\eta_3\eta_4} x_\gamma \, \sinh X_{\eta_2\eta_3\eta_4} }{ (X_{\eta_2\eta_3\eta_4}^2+x_\gamma^2)^2 \cosh x_1 \cosh x_2 \sinh x_3 \sinh x_4} \,.  \no\\
\label{eq:Fx1xgammaxcut}
\nq
It follows that
\bq
\Delta n_\ell &=& \int \frac{d^3 \bfk}{(2\pi)^3} L_{\bfk} \,, \no\\
F(x_\gamma, x_\text{cut}) &=& \int_0^{+\infty} x_1^2 dx_1 F(x_1, x_\gamma, x_\text{cut}), 
\nq
are satisfied and these results are compatible with our former work \cite{Pascoli:2016gkf}. 

The initial lepton asymmetry generated during the PT is not conserved but partially converted into the baryon asymmetry via the EW sphaleron processes which are unsuppressed above the EW scale. The $B-L$ asymmetry is a good symmetry and $n_{B-L} \equiv -\Delta n_L(T)$ is always conserved after the PT. The final baryon symmetry is 
 approximately given by $n_B \approx \frac{1}{3} n_{B-L}$. The baryon-to-photon ratio $\eta_B$ is defined as
\begin{eqnarray} 
\eta_B\equiv \frac{n_B}{n_\gamma} \approx \frac{\text{Im}\{\text{tr}[m_\nu^{0} m_\nu^*] \} T^2}{8 \pi^4 \zeta(3) v_H^4 } F(x_\gamma, x_\text{cut}),
\end{eqnarray}
where $n_\gamma= 2 \zeta(3) T^3/\pi^2$ with $\zeta (3) = 1.202$ have been used. 
In order to generate more baryon than anti-baryon, $\text{Im}\{\text{tr}[m_\nu^{0} m_\nu^*] \}$ should take a minus sign and it is worthnoting the lepton asymmetry is independent of the flavour basis we choose. A basis transformation $m_\nu^0\to \text{m}_\nu^0= U m_\nu^0 U^{T}$, $m_\nu\to \text{m}_\nu= U m_\nu U^{T}$ has no influence on the final lepton asymmetry since $\text{Im}\{\text{tr}[U\text{m}_\nu^{0} U^T U^* \text{m}_\nu^* U^\dag] \} = \text{Im}\{\text{tr}[m_\nu^{0} m_\nu^*] \}$ as expected.

\subsection{Numerical Analysis \label{sec:numerical}}

\begin{figure}[t!]
\centering
\includegraphics[width=0.75\textwidth]{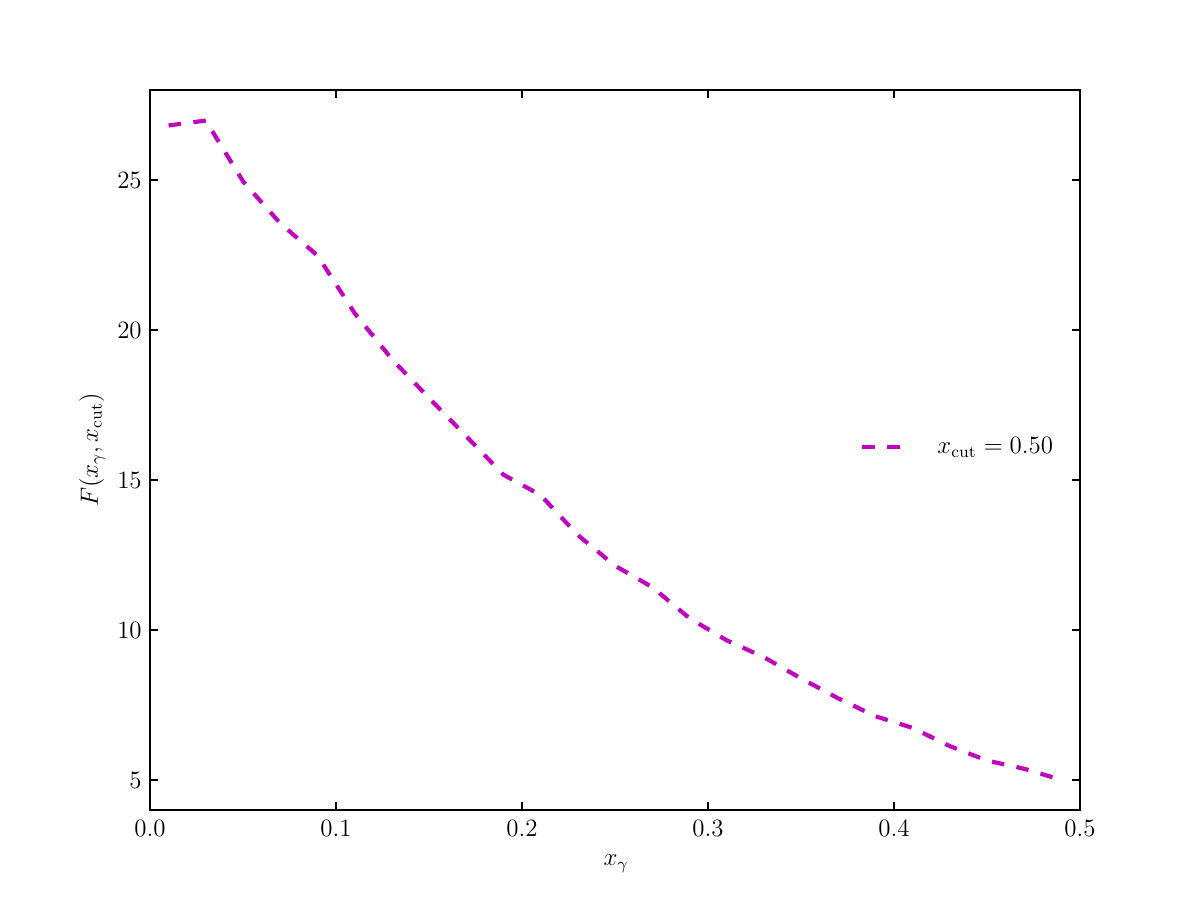}
\caption{\label{fig:loopfactor1}The loop factor $F(x_\gamma, x_\text{cut})$ as a function of $x_\gamma \equiv \gm/(2T)$, where $x_\text{cut}$ is fixed at $1/2$, corresponding to the energy transfer from the vacuum to the plasma being smaller than $T$. }
\end{figure}

The only factor which cannot be determined analytically is the loop factor $F(x_\gamma, x_\text{cut})$. In  \figref{fig:loopfactor1}, we fix $x_{\text{cut}}=1/2$  and show $F(x_\gamma, x_\text{cut})$ as a function of thermal width $x_\gamma \equiv \gm/(2T)$. Keeping in mind that $x_{\text{cut}}$ means the energy transfer from the vacuum to the Higgs and leptons less than $x_{\text{cut}} \times 2T$. $x_{\text{cut}}=1/2$ corresponds to the upper bound of energy transfer being $T$. For the phase transition at temperature $T$, it is natural to make such an assumption. The exact upper bound of the energy transfer may be different for this value. Indeed, we have varied $x_{\text{cut}}$ around 1/2, and found the integration $F(x_\gamma, x_\text{cut})$ is insensitive to the value $x_{\text{cut}}$. For $x_\gamma \sim \mathcal{O}(0.1)$ and $x_\text{cut} \sim \mathcal{O}(1)$, $F(x_\gamma, x_\text{cut})$ generally provides an $\mathcal{O}(10)$ factor enhancement. However, in some special PT, the energy transfer between the bubble wall and the plasma could be much smaller than the temperature. In that case, the value of the integration could be significantly suppressed and much smaller than $10$.

\begin{figure}[t!]
\centering
\includegraphics[width=0.75\textwidth]{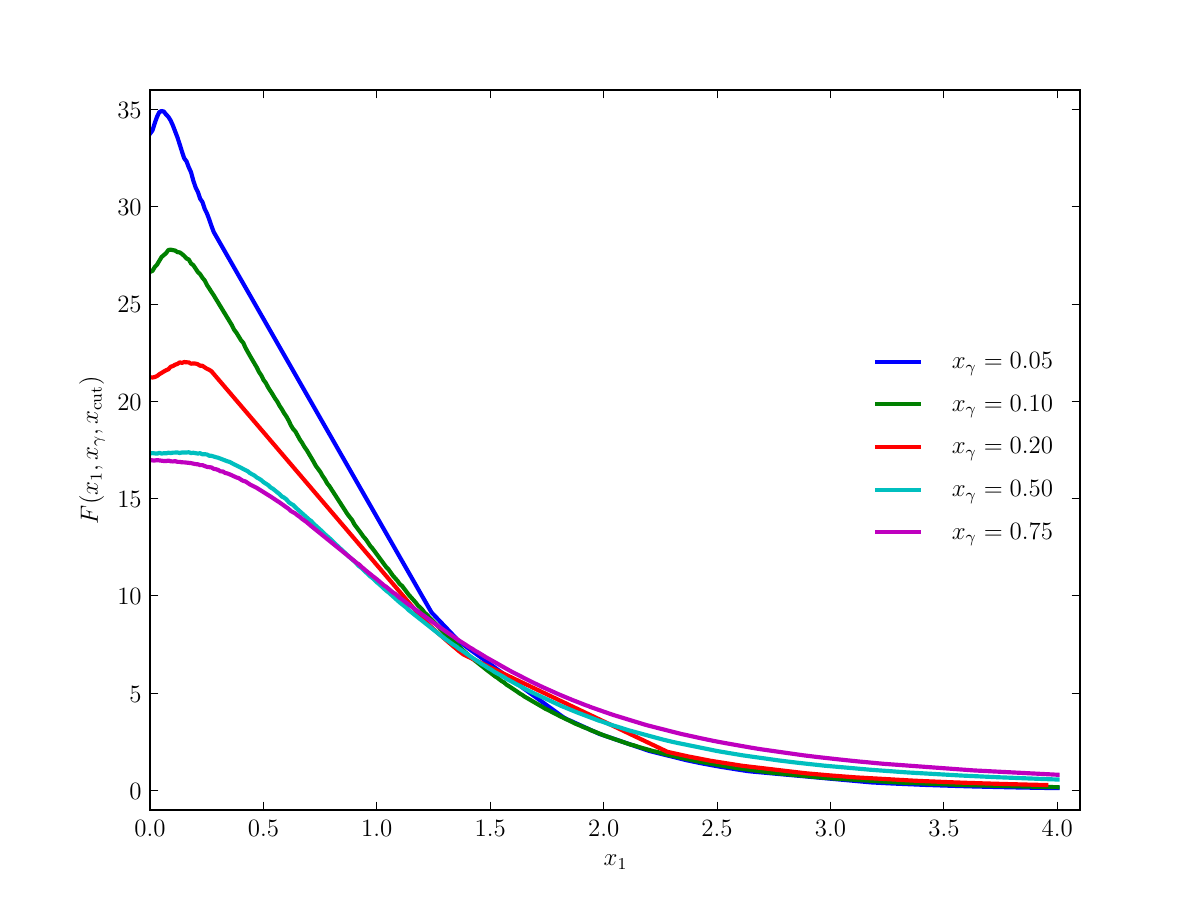}
\caption{\label{fig:loopfactor2}The loop factor $F(x_1, x_\gamma, x_\text{cut})$ as a function of $x_\gamma \equiv \gm/(2T)$, where $x_\text{cut}$ is fixed at 1/2, corresponding to the energy transfer from the vacuum to the plasma being smaller than $T$. }
\end{figure}

We show the momentum distribution $F(x_1,x_\gamma, x_\text{cut})$ as a function of $x_1$ with $x_\gamma$ fixed at $0.05, 0.1,0.2,0.5$ in  \figref{fig:loopfactor2}, respectively. In the Standard Model, the $x_\gamma\approx 0.1$, mostly originating from the contribution of EW gauge couplings \cite{bellac}. 

We estimate the temperature of successful leptogenesis. As discussed above, we can assume that the loop function $F(x_\gamma, x_\omega)$ provides an $\mathcal{O}(10)$ factor enhancement for $x_\gamma \sim \mathcal{O}(0.1)$ and $x_\omega \sim \mathcal{O}(1)$. Therefore, the final baryon asymmetry 
\bq 
\eta_B \sim \frac{\text{Im}\{\text{tr}[m_\nu^{0} m_\nu^*]\} T^2}{v_H^4} 10^{-2}  \,.
\nq
Since $\eta_B>0$, more baryon than anti-baryon, $\text{Im}\{\text{tr}[m_\nu^{0} m_\nu^*]\}$ must take a minus sign. 
In most cases, $\text{Im}\{\text{tr}[m_\nu^{0} m_\nu^*]\}$ is in the same order of $m_\nu^2$.
Then, we derive the PT temperature 
\bq
T \sim 10 \sqrt{\eta_B}\, \frac{v_H^2}{m_\nu} \approx 10^{11}\, \text{GeV} \,.
\nq

In our formalism, we do not consider the influence of temperature variation during the expansion of the Universe. This is valid if and only if the Hubble expansion rate $H$ is much smaller than the bubble wall expansion. The Hubble expansion rate is given by 
\bq
H = \frac{8 \pi}{3m_\text{pl}^2} = \frac{1.66 \sqrt{g_*}T^2}{m_\text{pl}},
\nq
where in the SM $g_*=106.75$; at such high scales it is possible  $g_*$ may be larger due to new degrees of freedom. Nevertheless, in general $g_*$ is a $\mathcal{O}(100)$ number and  therefore $H\sim \mathcal{O}(10) \sqrt{g_*}T^2/m_\text{pl}$. 
The bubble expansion rate is characterised by $v_w/L_w$ which is correlated with the bubble wall dynamics of the scalar $\phi^1$. 
To satisfy the requirement  $H\ll v_w/L_w$ assuming  $v_w/L_w \sim \mathcal{O}(0.01) T$, we find $T\ll \mathcal{O}(0.1) m_\text{pl}$ which is easily satisfied.

 The final result of the lepton asymmetry is crucially dependent upon the thermal width. In the limit $x_\gamma\to 0$, $F(x_1, x_\gamma)$  does not converge. This can be simply understood as follows. 
As previously discussed, CP violation is generated by the interference of two Weinberg operators at different times. 
To see more clearly where the divergence emerges, we consider a simplified case of the PT where the bubble wall is vanishingly thin:  $L_w\to 0$. Thus, given a fixed spatial point, the coefficient behaves as a step function along time where the Weinberg operators have steady coefficients $\lambda^0$ and $\lambda^0+\lambda^1$ before and after the PT respectively and the functional form of the coefficient is given by: $\lambda(t) = \lambda^0 + \lambda^1 \theta(t-t_0)$. Any interference between the Weinberg operator at time $t_1<t_0$ and  $t_2>t_0$ may generate a CP asymmetry no matter how large  the time difference, $|y|=|t_1-t_2|$. The thermal damping width corresponds to the decoherence effect of the Weinberg operator at a large time difference. In other words, as the thermal width becomes smaller, interference for larger $|y|$ will become increasingly significant.
In the limit of a zero-valued thermal width, interference between Weinberg operators in the infinite past and infinite future can also generate a lepton asymmetry, in addition to the lepton asymmetry generated at very short time differences. The size of the generated lepton asymmetry is almost the same but differs by a phase of the time difference. The total lepton asymmetry is obtained by the summation along time difference $y$ from $0$ to $\infty$, which does not converge but rather oscillates with $y$.
    Alternatively, one may consider the two-loop diagram of \figref{fig:Feynman}
as a self-energy correction to the lepton propagator. As the damping rate  is proportional to the imaginary component of the self-energy correction, taking the unphysical zero damping rate limit implies the two-loop correction vanishes and hence no lepton asymmetry is produced. 
We would like to emphasise that the treatment of the thermal widths we applied throughout this work constitute an effective treatment as the imaginary part of self-energy at finite temperature
is infrared divergent and gauge-dependent.  Generally,  one has to consider
gauge-field loops that generate the width explicitly, along with other
possible diagrams at the same order in the SM coupling and we relegate this particular issue for future study.

\section{Conclusion}\label{sec:final}

In this work we have provided a detailed discussion of leptogenesis via a varying Weinberg operator. The Weinberg operator violates  lepton number and $B-L$, which triggers processes of lepton-antilepton transition, di-lepton/di-antilepton annihilation and di-lepton/di-antilepton production. 
Motivated by tiny neutrino masses, the Weinberg operator is very weakly coupled. Thus, the triggered processes are slow and cannot reach thermal equilibrium for temperature below $10^{13}$ GeV. 
The spacetime variation of the Weinberg operator is fulfilled by including a CP-violating phase transition (CPPT). 

The novelties of this mechanism are: 
\begin{enumerate}
\item The realisation that the very weakly coupled Weinberg operator can fulfil the out-of-equilibrium condition.
\item The lepton asymmetry is generated via a phase transition and not via the decay of heavy particles. 
Consequently, a unique feature of the mechanism is the independence from a specific neutrino mass model because  all heavy particles have decoupled from the plasma before the phase transition. Therefore the Weinberg operator, obtained after all heavy particles are integrated out, is the only interaction violating $B-L$. The weakness of this operator also leads to the tiny washout effect which can be safely neglected. 
\end{enumerate}

In this paper we have presented the calculation of the lepton asymmetry from first principles, i.e., in the framework of non-equilibrium quantum field theory. Our calculation is entirely based on Green's functions. Such an approach avoids the need to separately calculate relevant processes as in the case of semi-classical Boltzmann equations. Our starting point was 
 a non-homogeneous scalar background in the rest plasma frame where we obtained the general expression of the lepton asymmetry in terms of the Wightman functions in the closed-time-path formalism. 

The feebly coupled Weinberg operator allowed us to analytically obtain the result of lepton asymmetry without considering time evolution. A non-zero lepton asymmetry is generated from the interference of spacetime-dependent Weinberg operators at different times. We provided an in depth  derivation of the lepton asymmetry generated by the varying Weinberg operator. In our calculation, two main contributions are specified: the dynamics of the PT and the thermal properties of the Higgs and leptons. 
We demonstrated the  lepton asymmetry factorises into a part proportional to the time-dependent coupling (the prefactor) and another part which involves integrating the finite-temperature matrix element over phase space. 

Although the nature of the PT does not alter the mechanism qualitatively, it influences the lepton asymmetry quantitatively. The contribution of the PT dynamics is represented as EEV profiles of some scalars $\langle \phi_i(x) \rangle$. The spacetime-varying coupling of the Weinberg operator is further represented as 
\bq
\lambda_{\alpha\beta}(x) = \lambda^0_{\alpha\beta} + \sum_{i=1}^n \lambda^{i}_{\alpha\beta}   \frac{\langle \phi_i(x) \rangle}{v_{\phi_i}} + \sum_{i,j=1}^n \lambda^{ij}_{\alpha\beta}  \frac{\langle \phi_i(x) \phi_j(x) \rangle}{v_{\phi_i} v_{\phi_j}} + \cdots \,. 
\nq 
These scalars may have complicated contributions to the final lepton asymmetry.
To simplify the discussion, we calculated the lepton asymmetry in the simplest sceanrio, the single scalar case where the coupling is represented as $\lambda_{\alpha\beta}(x) =  \lambda^0_{\alpha\beta} + \lambda^{1}_{\alpha\beta} {\langle \phi_i(x) \rangle} / {v_{\phi_i}}$. To evaluate the prefactor, we changed variables from times $t_1$, $t_2$ to the relative and average coordinate $r = x_2 -x_1$ and $x = (x_1 + x_2) /2$ and completed the spacetime integration. In the rest plasma frame, we separated the time and spatial integrations and proved that the latter is negligible. Therefore, the lepton asymmetry is mainly generated via the interference of Weinberg operator at different times. 

We discovered the connection between lepton asymmetry with neutrino masses, $\Delta n_L \propto \text{Im}\{\text{tr}[m_\nu^{0} m_\nu^*]$, where $m^0_\nu$ is the initial neutrino matrix before CPPT and $m_\nu$ is identical to the neutrino mass matrix we are to measure in neutrino experiments (ignoring RG effect running  from the scale $\Lambda_{\text{CPPT}}$ which have been shown to be small \cite{Casas:1999tg,Antusch:2003kp}). We also considered non-standard properties of the bubble, such as a slow-moving bubble with a thick wall, and the implications for this mechanism. However, we relegate a more extensive study of such cases for future work. 

Thermal properties of the Higgs and leptons, in particular their damping rates, are important. 
The interference of two Weinberg operators is dependent upon these damping rates. In order to generate a non-zero lepton asymmetry, the energy transfer between the leptons, Higgs and background must not conserved. This is unsurprising because there is a net energy transfer from the bubble wall to the Higgs and leptons. 

We have estimated the temperatures for successful leptogenesis. At high temperatures, the reaction rate of Weinberg operator is enhanced by $T^3$. Although this rate is small it is still sufficient to generate enough baryon yield for a given temperature. By assuming the prefactor of the same order of the neutrino mass, i.e., $\text{Im}\{\text{tr}[m_\nu^{0} m_\nu^*] \} \sim 0.1\, \text{eV}^2$ and the damping rates of the Higgs and leptons are approximately their SM values, we obtain that the phase transition at temperature $T_{\text{CPPT}} \sim 10^{11}$ GeV can generate $n_B \sim 10^{-10} n_\gamma$.

Compared with the well-known  EWBG, the PT in our mechanism plays a very different role.
While the PT is essential to generate the non-equilibrium state in EWBG, the Weinberg operator plays the key role in the departure of equilibrium in our mechanism. Such differing dynamics leads to many differences in the calculation and features of the final results, e.g., the spatial-independence in the integration in the rest plasma frame and the requirement of types of the PT, etc. However, these two mechanisms shares one similarity: the CP violation is generated by the PT. 

Finally, we comment that a first order PT has been assumed to simplify the calculation, although it is not a necessary condition to generate lepton asymmetry in the mechanism. If the PT is first ordered in nature, bubbles of the true vacuum nucleate and expand amongst the sea of the metastable phase in the universe. These bubbles finally meet and collide with each other giving rise to a significant stochastic background of gravitational waves \cite{Kosowsky:1992rz, Kamionkowski:1993fg}. This background resides today with the spectral shape peaked at a frequency related to the temperature of the PT. While 
eLISA \cite{Seoane:2013qna} will be capable of measuring EW-scale PT \cite{Caprini:2015zlo}, LIGO, Virgo and KAGRA has the potential to probe  PT for higher  temperatures  $\sim10^{7}-10^{12}$ GeV \cite{Abbott:2009ws, Aasi:2013wya, TheLIGOScientific:2016dpb}.

\acknowledgments
We would like to particularly thank  Silvia Pascoli for intensive discussions regarding this work. We are  also grateful to S. Bruggisser, M. Drewes, B. Garbrecht, B. T. Hambye, B. von Harling, P. Hern\'ardez, T. Konstandin, A. Long, S. Petcov, C. Tamarit, and D. Teresi for their useful discussions. 
This manuscript of  Y.L.Z. was supported by European Research Council under ERC Grant NuMass (FP7-IDEAS-ERC ERC-CG 617143),
and that of J.T. was authorised by Fermi Research Alliance, LLC under Contract No. DE-AC02-07CH11359 with the U.S. Department of Energy, Office of Science, Office of High Energy Physics. J.T. would like to express a special thanks to the Mainz Institute for Theoretical Physics (MITP) for their hospitality and support where part of this work was completed.
\appendix

\section{Examples of the EEV Profile}\label{sec:EEVprofle}
The exact expression for $\lambda(x)$ as a function of $x$ is determined by the properties of the PT. 
Here we introduce some specific types of profiles for $\lambda(x)$ in the bubble wall:  
\begin{itemize}
\item \textbf{Linear profile}, where $f(x')$ linear changes from 0 to 1 for $x'$ varying from 0 to $L_w$: 
\bq
f(x')= \left\{\begin{array}{ll}0\,, & x'<0 \\
{x'}/{L_w}\,, & 0<x'<L_w \\
1\,, & x'>L_w \end{array}\right.\,.
\nq
From this simple case, we can obtain steady spatial gradient of $\lambda$, $\partial_3 \lambda =\lambda^1/L_w$.  We note that 
a sudden change the scalar VEV can be triggered by dynamics other than a first order PT such as a quench in the context of cold EWBG \cite{Krauss:1999ng,GarciaBellido:1999sv}.

\item \textbf{Hyperbolic profile}, where the $\phi$ VEV takes the form of a hyperbolic function: 
\bq
f(x')= \frac{1}{2} \left[ 1+  \tanh\Big(\frac{x'}{L_w} \Big) \right] \,.
\nq
This case has been widely used as a numerical approximation of the Higgs VEV in EWBG \cite{Moreno:1998bq}. 
\end{itemize}
In the thin wall limit, ignoring the thickness of the bubble wall, i.e., $L_w\to 0$, we arrive at a Heaviside step function in both cases.

%%%%%%%%%%%%%%%%%%%%%%%%%%%%%%%%%%%%%%%%%%%%%%%%%%%%%%%%%%%%%%%%%%%%%%%%%%%%%%%%%%%%%%%%%%%%%%%%%%%%%%%%%%%%%%%%%%%%%%%%%%%%%%%%%%%%%%%%%%%%%%%%%%%%%%%%%%%%%%%%%%%%%

\section{Extensive Discussion on the Role of the Phase Transition}\label{sec:PT2}

In the main text, we calculated the lepton asymmetry with the assumption of a single scalar involved in the phase transition. Now we shall generalise this discussion to the multi-scalar case. Such an extension is necessary because many neutrino mass or flavour models involve more than one scalar. Multi-scalar phase transitions are the widely discussed in the context of the EW phase transition, which usually assumes additional scalar involving with the Higgs during the phase transition. 
Although a phase transition is necessary in CPPT  to generate the matter-antimatter asymmetry, the phase transition plays a very different role here in comparison with EWBG.
In the following, we will first discuss how the conclusion will be modified once extended to the multiple scalar case.

We extend our discussion to the two-scalar case. Ignoring the cross coupling between two scalars, the coupling matrix $\lambda(x)$ taking the following form 
\bq
\lambda(x)=\lambda^0+\lambda^1 f_1(x') + \lambda^2 f_2(x') \,,
\nq
Here, $f_1(x')$ and $f_2(x')$ correspond to EEV shapes of $\phi_1$ and $\phi_2$ respectively with $f_1(-\infty) = f_2(-\infty) =0$ and $f_1(+\infty) = f_2(+\infty) =1$. 
It is important that $\lambda^2$ takes a different relative phase compare with $\lambda^1$ 
and $f_2$ has a different profile from $f_1$. Otherwise, $\lambda^2$ and $f_2$ may be redefined to absorb $\lambda^1$ and $f_1$ respectively. With this consideration,
$\text{Im}\left\{ \text{tr} \left[\lambda^*(x_1) \lambda(x_2)\right] \right\}$ is simplified to 
\bq
\text{Im}\left\{ \text{tr} \left[\lambda^*(x_1) \lambda(x_2)\right] \right\} &=& \text{Im}\left\{ \text{tr} \left[\lambda^0 \lambda^{1*}\right] \right\} [f_1(x_1') - f_1(x_2')] + \text{Im}\left\{ \text{tr} \left[\lambda^0 \lambda^{2*}\right] \right\} [f_2(x_1') - f_2(x_2')]\no\\
&+& \text{Im}\left\{ \text{tr} \left[\lambda^{1} \lambda^{2*}\right] \right\} [f_1(x_2')f_2(x_1') - f_1(x_1')f_2(x_2')] \,.
\label{eq:int_x2}
\nq
The first line on the RHS will finally gives the same contribution to the lepton asymmetry as that of the RHS of \equaref{eq:lepton_asymmetry_f1}, $\text{Im}\{ \text{tr} [ m_\nu^{0} m_\nu^* ] \}$, which is independent of the shapes $f_1(x')$ or $f_2(x')$. The second and third lines represents the interference between the two scalar EEV profiles. Therefore, the lepton asymmetry generated by this term depends on the EEV shapes. 

In the case of vanishing initial coupling of the Weinberg operator $\lambda^0=0$, the lepton asymmetry can only be generated from the interference term. Typical examples are $U(1)_{B-L}$ models, where the symmetry forbid the initial coupling $\lambda^0$. Therefore, one has to introduce at least two scalars to generate a non-zero $\Delta n_\ell$. 
It is a possibility that there are more scalar EEV varying during the phase transition. Typical examples are flavour models. The inclusion of additional scalars into the system does not qualitatively alter the discussion but complicates the interference term.
 A careful discussion of the scalar contribution is related to detailed properties of the model, i.e., which symmetry is introduced, how many copies of scalars are in the model, coupling textures in the Weinberg operator, etc. We leave the relevant interesting studies to our future work.

The interference terms usually have very complicated contributions. We discuss two simplified cases where the first example is the multi-step phase transition. In other words, there exists a point $x_0'$, $f_1(x')$ varies from 0 to 1 for $x'$ running from $-\infty$ to $x_0'$ and $f_2(x')$ varies from 0 to 1 for $x'$ running from $x_0'$ to $+\infty$. The second line contributes a term $\text{Im}\{ \text{tr} [ \lambda^{1} \lambda^{2*} ] \}$, and $m_\nu = (\lambda^0+\lambda^1+\lambda^2) v_H^2 / \Lambda$. 

A second example is  the thick wall limit where the   following expansion is applied 
\bq
\int d^4r \text{Im}\{ \text{tr} [ \lambda^{*}(x_1) \lambda(x_2) ] \} \cM &=& \int d^4r \text{Im}\{ \text{tr} [ \lambda^{*}(x+r/2) \lambda(x-r/2) ] \} \cM \nonumber\\
&\approx& \text{Im}\{ \text{tr} [ \lambda^{*}(x) \partial_\mu \lambda(x) ] \} \int d^4r r^\mu  \cM \,.
\nq
For $\mu=0$ and $\mu=3$, we get the time- and space-dependent lepton asymmetries. 
\bq
\Delta n^{\text{I}} &\propto& \text{Im}\{ \text{tr} [ \lambda^{*}(x) \partial_t\lambda (x) ] \} \equiv \sum_{\alpha\beta} |\lambda_{\alpha\beta}(x)|^2 \partial_t\phi_{\alpha\beta} (x)  \,, \no\\
\Delta n^{\text{II}} &\propto& \text{Im}\{ \text{tr} [ \lambda^{*}(x) \partial_z\lambda(x)] \} \equiv \sum_{\alpha\beta} |\lambda_{\alpha\beta}(x)|^2 \partial_z\phi_{\alpha\beta} (x)  \,, 
\nq
respectively where $t= x^0$ and $z=x^3$. The CP source of $\Delta n^{\text{II}}$ takes a similar form as that in EWBG, which is proportional to $\Imag\{ \tr[ m^*_q \partial_z m^T_q ] \}$, where $m_q$ is the quark mass matrix in the flavour space \cite{Prokopec:2003pj,Prokopec:2004ic}. 
At lower temperatures, where the deviation from thermal equilibrium grows, $\Delta n^{\text{II}}$ has an enhanced contribution. However, as we are considering temperatures much higher than the EW scale, where the equilibrium distributions for the Higgs and leptons are assumed in $\cM$, we find that the space-dependent lepton asymmetry is vanishing. Therefore, the total lepton asymmetry is proportional to
\bq
\Delta n_\ell &\propto& \frac{1}{v_H^4} \int_{-\infty}^{+\infty} dt \, \text{Im}\{ \text{tr} [ m_\nu^{*}(x) \partial_t m_\nu(x) ] \} \int d^4r y \,  \cM \no\\
&\propto& \frac{v_w}{v_H^4} \int_{-\infty}^{+\infty} dt \, \text{Im}\{ \text{tr} [ m_\nu^{*}(x) \partial_z m_\nu(x) ] \} \int d^4r y \, \cM,
\nq
where $m_\nu(x) \equiv \lambda(x) v_H^2/\Lambda$ and $\partial_0 = - v_w \partial_3$ have been used. 
It is useful to define the CP sources per unit volume per unit time $S_\ell(x)$ as 
\bq
S_\ell (x) &=& - \frac{12}{v_H^4} \text{Im}\{ \text{tr} [ m_\nu^{*}(x) \partial_t m_\nu(x) ] \} \int d^4r y  \cM \nonumber\\
&=& v_w \frac{12}{v_H^4} \text{Im}\{ \text{tr} [ m_\nu^{*}(x) \partial_z m_\nu(x) ] \} \int d^4r y  \cM \,.
\label{eq:source}
\nq 
Naively, we find
$n_L \approx \frac{L_w}{v_w} \overline{S}_\ell$, 
where $\overline{S}_\ell$ is the mean value of $S_\ell(x)$ in the wall. 
In our work, we assume the bubble expansion is sufficiently fast that the effect of Hubble expansion, i.e., the evolution with temparature/time, may be ignored. In the slow bubble expansion case that $L_w/v_w \gtrsim 1/H_u$, the effect of Hubble expansion should be included.

%%%%%%%%%%%%%%%%%%%%%%%%%%%%%%%%%%%%%%%%%%%%%%%%%%%%%%%
%%%%%%%%%%%%%%%%%%%%%%%%%%%%%%%%%%%%%%%%%%%%%%%%%%%%%%%
%%%%%%%%%%%%%%%%%%%%%%%%%%%%%%%%%%%%%%%%%%%%%%%%%%%%%%%
%%%%%%%%%%%%%%%%%%%%%%%%%%%%%%%%%%%%%%%%%%%%%%%%%%%%%%%

\section{Matrix Element}\label{sec:ME}

In this Appendix, we provide some additional details on the calculation of the  matrix element.
It may be shown that the  matrix element, $M$, of \equaref{eq:traceexp} may be rewritten such that
\begin{equation}\label{eq:ME}
M = \text{Im}\big\{ \Delta^{<}_{\bfq}(t_1,t_2)  \Delta^{<}_{\bfq^{\prime}}(t_1,t_2) \text{tr} \big[ S^{<}_{\bfk}(t_1,t_2)S^{<}_{\bfk^{\prime}}(t_1,t_2) P_{L}  \big]   \big\}.
\end{equation}
We apply the CTP Feynman rules and use the free equilibrium propagators of the massless leptons and Higgs field which are given by \cite{Anisimov:2010dk} 
\begin{eqnarray}
    \Delta_{\bfq}^{<}\left(y \right)&=&\frac{1}{2 \omega_{\bfq}}\left[\coth\left(\frac{\beta\omega_{\bfq}}{2}\right) \cos\left( \omega_{\bfq}y\right)+i\sin(  \omega_{\bfq}y) \right], \\
 S_{\bfk}^{<}\left(y \right)&=&  -\frac{\gamma^0}{2}\left[\cos\left( \omega_{\bfk}y\right) + i \tanh\left(\frac{\beta \omega_{\bfk}}{2}\right)\sin\left( \omega_{\bfk}y\right)\right] \no\\
 &&- \frac{\vec{\gamma}\cdot\bfk}{2\omega_{\bfk}}\left[ \tanh\left(\frac{\beta \omega_{\bfk}}{2}\right)\cos\left( \omega_{\bfk}y\right)+i\sin\left( \omega_{\bfk}y\right) \right],
\end{eqnarray}
where  $\beta=1/T$ and have applied the notation of the relative coordinate, $y$, for brevity. These propagators may be simplified using the redefinition of the relative coordinate $y^-=y-i\beta/2$ 
\bq
\label{eq:props}
\Delta_{\bfq}^{<}\left(y \right)
&=& \frac{1}{2  \omega_{\bfq}}\left[ \coth\left(\frac{\beta  \omega_{\bfq}}{2}\right) \cos\left(  \omega_{\bfq}y^-+\frac{i\beta  \omega_{\bfq}}{2}\right)+i \sin\left(  \omega_{\bfq}y^-+\frac{i  \omega_{\bfq}\beta}{2}\right) \right] \no\\
&=& \frac{1}{2  \omega_{\bfq}}\frac{ \cos\left(  \omega_{\bfq}y^-\right)}{\sinh\left(\frac{  \omega_{\bfq}\beta}{2}\right)}\,, \no\\
S_{\bfk}^{<}\left(y \right) &=& - \frac{\gamma^{0}}{2}\left[ \cos\left( \omega_{\bfk}y^-+\frac{i\beta \omega_{\bfk}}{2}\right) + i \tanh\left(\frac{\beta \omega_{\bfk}}{2}\right)\sin\left( \omega_{\bfk}y^-+\frac{i\beta \omega_{\bfk}}{2}\right)\right]\no\\
&&-\frac{\vec{\gamma}\cdot\hat{\bfk}}{2}\left[ \tanh\left(\frac{\beta \omega_{\bfk}}{2}\right)\cos\left( \omega_{\bfk}y^-+\frac{i\beta \omega_{\bfk}}{2}\right)+i\sin\left( \omega_{\bfk}y^-+\frac{i\beta \omega_{\bfk}}{2}\right)  \right] \no\\
&=& -\frac{\gamma^{0}\cos\left( \omega_{\bfk}y^-\right)+i\vec{\gamma}\cdot\hat{\bfk} \sin\left( \omega_{\bfk}y^-\right)}{2\cosh\left(\frac{\beta \omega_{\bfk}}{2}\right)}\,,
\nq
where we have  applied the notation  $\hat{\bfk}=\bfk/\omega_\bfk$. Naturally, for left-handed fermions $S_{\bfk}^{<}\rightarrow P_LS_{\bfk}^{<}$.   Multiplying out these propagators we find
\bq
\label{eq:DDSS}
&&\Delta_{\bfq^{\prime}}^{<}\Delta_{\bfq}^{<}\text{tr}\Big[S_{\bfk}^{<}S_{\bfk^{\prime}}^{<}\Big]=\frac{\cos\left( \omega_{\bfq}y^-\right)\cos\left( \omega_{\bfq^{\prime}}y^-\right)}{4 \omega_{\bfq} \omega_{\bfq^{\prime}}\sinh\left(\frac{ \omega_{\bfq}\beta}{2}\right)\sinh\left(\frac{ \omega_{\bfq^{\prime}}\beta}{2}\right)} \no\\
&&\times\left[\text{Tr}\left(\frac{\gamma^{0}\gamma^{0}}{4}\right)\frac{\cos\left( \omega_{\bfk}y^-\right)\cos\left( \omega_{\bfk^{\prime}}y^-\right)}{\cosh\left(\frac{  \omega_{\bfk}\beta}{2}\right)\cosh\left(\frac{ \omega_{\bfk^{\prime}}\beta}{2}\right)} -\text{Tr}\left(\frac{\gamma^i\gamma^j}{4}\right)\frac{\hat{\bfk}_i\hat{\bfk}_j^{\prime}\sin\left(  \omega_{\bfk}y^-\right)\sin\left( \omega_{\bfk^{\prime}}y^-\right)}{\cosh\left(\frac{  \omega_{\bfk}\beta}{2}\right)\cosh\left(\frac{ \omega_{\bfk^{\prime}}\beta}{2}\right)}   \right] \no\\
&&=\frac{1}{8 \omega_{\bfq} \omega_{\bfq^{\prime}}} \frac{\cos\left( \omega_{\bfq}y^-\right)
\cos\left( \omega_{\bfq^{\prime}}y^-\right)}{\sinh\left(\frac{ \omega_{\bfq}\beta}{2}\right)\sinh\left(\frac{ \omega_{\bfq^{\prime}}\beta}{2}\right)\cosh\left(\frac{ \omega_{\bfk}\beta}{2}\right)\cosh\left(\frac{\omega_{\bfk^{\prime}}\beta}{2}\right)} \no\\
&&\times\left(\cos\left( \omega_{\bfk}y^-\right)\cos\left( \omega_{\bfk^{\prime}}y^-\right)+\delta^{ij}\hat{\bfk}_{i}\hat{\bfk}^\prime_{j}\sin\left( \omega_{\bfk}y^-\right)\sin\left( \omega_{\bfk^{\prime}}y^-\right)\right) \,.
\nq
Taking the imaginary part and appending the above with the appropriate thermal damping rates ($e^{-\gm\lvert y \rvert}$), we recover \equaref{eq:eqprop}.
The matrix element can be further expanded and to do so  we denote the numerator of 
$\text{tr}\Big[S_{\bfk}^{<}S_{\bfk^{\prime}}^{<}\Big]\Delta_{\bfq^{\prime}}^{<}\Delta_{\bfq}^{<}$ as 
\bq
\label{eq:notsimp}
\underbrace{\cos\left(  \omega_{\bfq}y^-\right)\cos\left( \omega_{\bfq^{\prime}}y^-\right)}_{f_{1}}  \Big[\underbrace{\cos\left( \omega_{\bfk}y^-\right)\cos\left( \omega_{\bfk^{\prime}}y^-\right)}_{f_{2}} \!+\!\underbrace{\delta^{ij}\hat{\bfk}_{i}\hat{\bfk}^\prime_{j} \sin\left(  \omega_{\bfk}y^-\right)\sin\left(\omega_{\bfk^{\prime}}y^-\right)}_{f_{3}}\Big]\,. 
\nq
Multiplying out $f_1\times f_2$ we find
\bq
\label{eq:simp1}
f_{1} \times f_{2} &=&\left[ \frac{e^{i\left( \omega_{\bfq}+ \omega_{\bfq^{\prime}}\right)y^-}+e^{i\left( \omega_{\bfq}- \omega_{\bfq^{\prime}}\right)y^-}+e^{i\left( \omega_{\bfq^{\prime}}- \omega_{\bfq}\right)y^-}+e^{-i\left( \omega_{\bfq}+ \omega_{\bfq^{\prime}}\right)y^-}}{4}\right] \no\\
&&\times \left[ \frac{e^{i\left( \omega_{\bfk}+ \omega_{\bfk^{\prime}}\right)y^-}+e^{i\left( \omega_{\bfk}- \omega_{\bfk^{\prime}}\right)y^-}+e^{i\left( \omega_{\bfk^{\prime}}- \omega_{\bfk}\right)y^-}+e^{-i\left( \omega_{\bfk}+ \omega_{\bfk^{\prime}}\right)y^-}}{4}\right]  \no\\
&=&\frac{1}{16} \Big[e^{i\left(  \omega_{\bfq}+ \omega_{\bfq^{\prime}} +  \omega_{\bfk}+ \omega_{\bfk^{\prime}}  \right)y^-} + e^{i\left(  \omega_{\bfq}- \omega_{\bfq^{\prime}} +  \omega_{\bfk}+ \omega_{\bfk^{\prime}}  \right)y^-} +
e^{i\left( - \omega_{\bfq}+ \omega_{\bfq^{\prime}} +  \omega_{\bfk}+ \omega_{\bfk^{\prime}}  \right)y^-} \no\\ 
&& + e^{i\left( - \omega_{\bfq}- \omega_{\bfq^{\prime}} +  \omega_{\bfk}+ \omega_{\bfk^{\prime}}  \right)y^-} 
+ e^{i\left(  \omega_{\bfq}+ \omega_{\bfq^{\prime}} +  \omega_{\bfk}- \omega_{\bfk^{\prime}}  \right)y^-}  + e^{i\left(  \omega_{\bfq}- \omega_{\bfq^{\prime}} +  \omega_{\bfk}- \omega_{\bfk^{\prime}}  \right)y^-} \no\\
&&+ e^{i\left( - \omega_{\bfq}+ \omega_{\bfq^{\prime}} +  \omega_{\bfk}- \omega_{\bfk^{\prime}}  \right)y^-} 
 + e^{i\left( - \omega_{\bfq}- \omega_{\bfq^{\prime}} +  \omega_{\bfk}- \omega_{\bfk^{\prime}}  \right)y^-}\Big]+ \text{c.c.}\,.
\nq
Recalling $y^- = y-i\beta/2$, we may make the expansion  $e^{i\left(xy-i x\beta/2\right)} \equiv e^{ixy}e^{\beta x/2}$. To find the imaginary part this implies $\text{Im}[e^{ixy}e^{\beta x/2}]\equiv \sin(xy)e^{\beta x/2}$. 
Applying this to \equaref{eq:simp1} we find
\bq
\label{eq:simp2}
f_{1} \times f_{2} &=& \frac{1}{16} \Big[ \sin\left( \omega_{\bfq}+ \omega_{\bfq^\prime} +  \omega_{\bfk}+ \omega_{\bfk^{\prime}}  \right) e^{\beta/2\left( \omega_{\bfq}+ \omega_{\bfq^\prime} +  \omega_{\bfk}+ \omega_{\bfk^{\prime}} \right)} \no\\
&&~~ +\sin\left( \omega_{\bfq}- \omega_{\bfq^\prime} +  \omega_{\bfk}+ \omega_{\bfk^{\prime}}  \right)e^{\beta/2\left( \omega_{\bfq}- \omega_{\bfq^\prime} +  \omega_{\bfk}+ \omega_{\bfk^{\prime}} \right)} \no\\
&&~~ +\sin\left(- \omega_{\bfq}+ \omega_{\bfq^\prime} +  \omega_{\bfk}+ \omega_{\bfk^{\prime}}  \right)e^{\beta/2\left(- \omega_{\bfq}+ \omega_{\bfq^\prime} +  \omega_{\bfk}+ \omega_{\bfk^{\prime}}  \right)} \no\\
&&~~ +\sin\left(- \omega_{\bfq}- \omega_{\bfq^\prime} +  \omega_{\bfk}+ \omega_{\bfk^{\prime}}   \right)e^{\beta/2\left(- \omega_{\bfq}- \omega_{\bfq^\prime} +  \omega_{\bfk}+ \omega_{\bfk^{\prime}}   \right)} \no\\
&&~~ +\sin\left( \omega_{\bfq}+ \omega_{\bfq^\prime} +  \omega_{\bfk}- \omega_{\bfk^{\prime}} \right)e^{\beta/2\left( \omega_{\bfq}+ \omega_{\bfq^\prime} +  \omega_{\bfk}- \omega_{\bfk^{\prime}}\right)} \no\\
&&~~ +\sin\left( \omega_{\bfq}- \omega_{\bfq^\prime} +  \omega_{\bfk}- \omega_{\bfk^{\prime}} \right)e^{\beta/2\left( \omega_{\bfq}- \omega_{\bfq^\prime} +  \omega_{\bfk}- \omega_{\bfk^{\prime}}\right)} \no\\
&&~~ +\sin\left(- \omega_{\bfq}+ \omega_{\bfq^\prime} +  \omega_{\bfk}- \omega_{\bfk^{\prime}}  \right)e^{\beta/2\left(- \omega_{\bfq}+ \omega_{\bfq^\prime} +  \omega_{\bfk}- \omega_{\bfk^{\prime}} \right)} \no\\
&&~~ +\sin\left(- \omega_{\bfq}- \omega_{\bfq^\prime} +  \omega_{\bfk}- \omega_{\bfk^{\prime}} \right)e^{\beta/2\left(- \omega_{\bfq}- \omega_{\bfq^\prime} +  \omega_{\bfk}- \omega_{\bfk^{\prime}}\right)} \Big] + \text{c.c.} \,.
\nq
The complex conjugate from above is treated in the following way: $e^{-ixy^-}\equiv e^{-i(xy-i\beta x/2)}=e^{-ixy}e^{-\beta x/2}\implies \text{Im}[e^{-ixy}e^{-\beta x/2}]=-\sin\left(xy\right)e^{-\beta x/2}$.
There adding to its complex conjugate, we find $\sin\left(xy\right)e^{\beta x/2}-\sin\left(xy\right)e^{-\beta x/2}=2\sin\left(xy\right)\sinh\left(\beta x/2\right)$. This implies \equaref{eq:simp1} may be written as
\bq
\label{eq:simp3}
f_{1} \times f_{2} = \frac{2}{16}\!\!\! &&\Bigg[
\sin\left(K_{+++}y\right)\sinh\left(\frac{\beta K_{+++}}{2}\right) + \sin\left(K_{++-}y\right)\sinh\left(\frac{\beta K_{++-}}{2}\right) \no\\
&&+\sin\left(K_{+-+}y\right)\sinh\left(\frac{\beta K_{+-+}}{2}\right) + \sin\left(K_{+--}y\right)\sinh\left(\frac{\beta K_{+--}}{2}\right) \no\\
&&+\sin\left(K_{-++}y\right)\sinh\left(\frac{\beta K_{-++}}{2}\right) + \sin\left(K_{-+-}y\right)\sinh\left(\frac{\beta K_{-+-}}{2}\right) \no\\
&&+\sin\left(K_{--+}y\right)\sinh\left(\frac{\beta K_{--+}}{2}\right) + \sin\left(K_{---}y\right)\sinh\left(\frac{\beta K_{---}}{2}\right]  \Bigg\}\,,
\nq
where we have applied the following definitions for ease of notation
\bq
\label{eq:simp4}
K_{+++}&= \omega_{\bfk}+ \omega_{\bfk^{\prime}}+ \omega_{\bfq}+ \omega_{\bfq^\prime}\,, \no\\
K_{++-}&= \omega_{\bfk}+ \omega_{\bfk^{\prime}}+ \omega_{\bfq}- \omega_{\bfq^\prime}\,, \no\\
K_{+-+}&= \omega_{\bfk}+ \omega_{\bfk^{\prime}}- \omega_{\bfq}+ \omega_{\bfq^\prime}\,, \no\\
K_{+--}&= \omega_{\bfk}+ \omega_{\bfk^{\prime}}- \omega_{\bfq}- \omega_{\bfq^\prime}\,, \no\\
K_{-++}&= \omega_{\bfk}- \omega_{\bfk^{\prime}}+ \omega_{\bfq}+ \omega_{\bfq^\prime}\,, \no\\
K_{-+-}&= \omega_{\bfk}- \omega_{\bfk^{\prime}}+ \omega_{\bfq}- \omega_{\bfq^\prime}\,, \no\\
K_{--+}&= \omega_{\bfk}- \omega_{\bfk^{\prime}}- \omega_{\bfq}+ \omega_{\bfq^\prime}\,, \no\\
K_{---}&= \omega_{\bfk}- \omega_{\bfk^{\prime}}- \omega_{\bfq}- \omega_{\bfq^\prime}\,,
\nq
where  $K_{\eta_{2}\eta_{3}\eta_{4}}= \omega_\bfk+\eta_{2} \omega_{\bfk^{\prime}}+\eta_{3} \omega_\bfq+\eta_{4} \omega_{\bfq^{\prime}}$ and $\eta_{i}=\pm 1$ for $i= 1,2,3$.
Applying  the same procedure, we  calculate $f_1\times f_3$
\bq
\label{eq:simp5}
f_{1}\times f_{3}  = \frac{\hat{\bfk} \cdot \hat{\bfk}^{\prime}}{16} \Big[  
\!\!\!&&-\sin\left( K_{+++}y\right)e^{\beta K_{+++}/2} -\sin\left(K_{++-}y\right)e^{\beta K_{++-}/2} \no\\
&&-\sin\left(K_{+-+}y\right)e^{\beta K_{+-+}/2} -\sin\left(K_{+--}y\right)e^{\beta K_{+--}/2} \no\\
&&+\sin\left(K_{-++}y\right)e^{\beta K_{-++}/2} +\sin\left(K_{-+-}y\right)e^{\beta K_{-+-}/2} \no\\
&&-\sin\left(K_{--+}y\right)e^{\beta K_{--+}/2} +\sin\left(K_{---}y\right)e^{\beta K_{---}/2} +\text{c.c.}
\Big]\,.
\nq
Adding the complex conjugate part in the same way as before we find
\bq
\label{eq:simp6}
\hspace{-1cm}f_{1}\times f_{3}  = \frac{2\hat{\bfk} \cdot \hat{\bfk}^{\prime}}{16} \Bigg[  
\!\!\!&& -\sin\left( K_{+++}y \right)\sinh \left(\frac{\beta K_{+++}}{2}\right) -\sin\left( K_{++-}y \right)\sinh \left(\frac{\beta K_{++-}}{2}\right)\no\\
&& -\sin\left( K_{+-+}y \right)\sinh \left(\frac{\beta K_{+-+}}{2}\right) -\sin\left( K_{+--}y \right)\sinh \left(\frac{\beta K_{+--}}{2}\right)\no\\
&& +\sin\left( K_{-++}y \right)\sinh \left(\frac{\beta K_{-++}}{2}\right) +\sin\left( K_{-+-}y \right)\sinh \left(\frac{\beta K_{-+-}}{2}\right)\no\\
&& +\sin\left( K_{--+}y \right)\sinh \left(\frac{\beta K_{--+}}{2}\right) +\sin\left( K_{---}y \right)\sinh \left(\frac{\beta K_{---}}{2}\right)\Bigg].
\nq
Collecting all the terms and using the following relation
\begin{equation} \label{eq:int_0_infty}
\int_{0}^{\infty} dy \, y \, \sin\left(K y \right)e^{-\gamma y} = \frac{2K \gamma}{\left(K^2+\gamma^2\right)^2},
\end{equation}
to complete the integration over $y$, we recover \equaref{eq:inteM}. 
 
It is worthwhile to note that this integration is only valid in the case of the finite width, namely $\gamma >0$.
In the limit $\gamma\to 0$, we encounter the oscillating problem of the integral. 
A physical interpretation of this behaviour has been given at  the end of \secref{sec:numerical}. 
To see this more clearly, we go back to the initial integration $\int d^4 x_1 d^4 x_2$ and replace the interval of the time component from \eqref{eq:int_0_infty} from $(-\infty,+\infty)$ to $[-t/2,t/2]$. We further follow the technique use in \cite{Anisimov:2010dk} (see the discussion from Eq. (5.41) and therein) by defining the integrals
\begin{eqnarray} \label{eq:int_It}
\widetilde{\mathcal{I}}(t) &=& \int_{-t/2}^{t/2} d t_1 \int_{-t/2}^{t/2} d t_2  \, (t_1-t_2) \,  e^{-i \Omega_1 t_1 + i\Omega_2 t_2} e^{-\gamma (t_1+t_2)} \,,
\end{eqnarray}
where $\Omega_i$ is a function of particle energies. It is useful to parameterise $\Omega_i = z_i \gamma$ for later use. 
While $\widetilde{\mathcal{I}}(t)$, which involves time-difference $t_1-t_2\equiv y$, is the key integral in our mechanism, $\mathcal{I}(t)$ as defined in \cite{Anisimov:2010dk} (see Eq.~(6.2)) does not include this factor. 
It has been rectified in  \cite{Anisimov:2010dk} (see Appendix E) that the CTP result of thermal leptogenesis can recover the Boltzmann result in the zero-width limit.  The main point is that the integral $\mathcal{I}(t) + \mathcal{I}^*(t)$, after integrating times, contains $(z_i^2+1)$ in the denominator. Thus $\mathcal{I}(t)$ has simple poles at $z_i = \pm i$. In the limit $\gamma/K \to 0$ with $\tau = \gamma t$ fixed, integration along $z_i$ is expanded to the interval $(-\infty, +\infty)$, which can be further spanned to the closed path encircling the whole upper complex plane or lower complex plane. In this case, the Cauchy's theorem applies and the integral convergences to a finite value. 
We are going to check the behaviour of our mechanism from the same mathematical point of view. 
We straightforwardly yield
\begin{eqnarray}
\widetilde{\mathcal{I}}(t) + \widetilde{\mathcal{I}}^*(t) &=& \frac{2 t^3 z_1^2 z_2^2 (\sin z_1-\sin z_2)-e^{-\tau}(z_1+z_2) \sin (z_1-z_2) + \cdots}{\left(z_1^2+1\right)^2 \left(z_2^2+1\right)^2} \,.
\end{eqnarray}
As the numerator is irrelevant for our discussion we do not provide the full expression. The most important feature we highlight here is that  $\widetilde{\mathcal{I}}(t)+\widetilde{\mathcal{I}}^*(t)$ has poles of order 2 at $z_i=\pm i$. Thus we cannot apply Cauchy's theorem and we encounter a divergence. 
We have also checked if $(t_1-t_2)$ in the integrand in Eq.~\eqref{eq:int_It} is abandoned, $z_i=\pm i$ turn to simple poles, and the integration along $z_i$ is finite.

%%%%%%%%%%%%%%%%%%%%%%%%%%%%%%%
%%%%    EWBG						%%%%%%%
%%%%%%%%%%%%%%%%%%%%%%%%%%%%%%%

\section{Comparison with EWBG}\label{sec:EWBG}

The best known mechanism of PT-induced baryogenesis is EWBG. 
Although our mechanism shares a common feature with EWBG, that being a PT driving the generation of the
baryon asymmetry,  CPPT differs greatly from EWBG.
The differences between these two mechanisms originate from how the three Sakharov conditions are satisfied. The essential differences are listed as follows.
\begin{itemize}
\item In EWBG, the baryon number violation is provided by sphaleron transitions in the symmetric phase. Both the out-of-equilibrium condition and C/CP violations are induced by EW phase transition \cite{Morrissey:2012aa}. In the EWBG, the phase transition is key to the generation of the non-equilibrium evolution. In order to achieve this, rapidly expanding bubble walls  are required such that the backreactions are not efficient to wash out the generated baryon asymmetry. 
\item As originally considered in Ref. \cite{Pascoli:2016gkf}, and further elucidated in \secref{sec:intro}, the $B-L$ number violation and departure from thermodynamic equilibrium are directly provided by the very weakly coupled Weinberg operator. 
The PT is only necessary to provide a source of  C/CP violation and is not needed for 
the efficiency of reactions in the system. Consequently,  successful leptogenesis in this setup does not  necessarily require a  first-order PT and it is possible a CP-violating second-order PT would also  generate a lepton asymmetry. The purpose of assuming the first-order phase transition in the former sections is to simplify the discussion and derive the lepton asymmetry quantitatively. 
\end{itemize}

With reference to the differing non-equilibrium dynamics provided in these two mechanisms, the method of  calculation varies. 
For example, in our mechanism it is not necessary to boost to the {\bf rest wall frame} as in the case of EWBG. In the rest frame of the wall, the particle distribution is not isotropic thus both the time-dependent and space-dependent integration will be non-zero. 
In EWBG, the non-isotropic component of the particle (e.g., the top quark and Higgs) distribution in front of the bubble wall is much larger. Thus, the space-dependent integration in the rest frame of the plasma may have a sizeable contribution to the baryon asymmetry. 

One may wonder to what extent the non-equilibrium distribution may give rise to a non-zero spatial-dependent integration and the subsequent contribution to the lepton asymmetry. To estimate this effect let us assume, in the rest frame of the plasma, there is a small non-isotropic deviation  the equilibrium for leptons, i.e., replacing $f_{F, |k^0|}(x)$ in Eq.~\eqref{eq:equilibrium} by 
\bq
f_{\ell, \bfk}(x) &=& f_{F, |k^0|}[1 + \epsilon_{\ell, \bfk} (x) + \cdots] \,,\nonumber\\
f_{\overline{\ell}, \bfk}(x) &=& f_{F, |k^0|}[1 + \epsilon_{\overline{\ell}, \bfk} (x) + \cdots] \,,
\nq
where the dots are irrelevant isotropic corrections. 
The bubble wall is the only source of non-isotropy and  influences the distribution of leptons only via the Weinberg operator
\bq
\epsilon_{\ell, \bfk}(x), \epsilon_{\overline{\ell}, \bfk} (x) \sim \frac{|\lambda|^2}{\Lambda^2} T^2 \,.
\nq 
Then, the spatial integration $\Delta n_\ell^{\text{II}} \sim \lambda^2 T^2 /\Lambda^2 \Delta n_\ell^{\text{I}} \ll \Delta n_\ell^{\text{I}}$. 

With reference the discussion shown there, the space integration in {\bf the rest frame of the plasma} is zero. This result is obtained from the assumption that the Higgs and leptons are almost in thermal equilibrium in the source term which is justifiable at such temperatures. While in the case of EWBG, charge separation induced by the Higgs may lead to non-negligible spatial distribution. 

Finally, we draw a comparison between the contribution of the PT in both mechanisms. To further elucidate, we assume a two-scalar phase transition with $\lambda^0 = 0$, $\lambda (x) = \lambda^1 \phi_1 / v_{\phi_1} + \lambda^2 \phi_2 / v_{\phi_2}$. Then, the CP source  is
\bq
S_\ell (x) = - \frac{12}{v_H^4} \text{Im}\{ \text{tr} [ m_\nu^{1*} m_\nu^{2} ] \} \times \left[ f_1(x') \partial_t f_2(x') - f_2(x') \partial_t f_1(x') \right] \times \int d^4r y  \cM.
\nq 
The middle term on the right hand side shows the dependence of the first derivative on the VEV profile. This property has been obtained in supersymmetric EWBG in the approximation of VEV insertion \cite{Carena:2000id,Lee:2004we}. 

%%%%%%%%%%%%%%%%%%%%%%%%%%%%%%%%%%%%%%%%%%%%%%%%%%%%%%%
%%%%%%%%%%%%%%%%%%%%%%%%%%%%%%%%%%%%%%%%%%%%%%%%%%%%%%%
%%%%%%%%%%%%%%%%%%%%%%%%%%%%%%%%%%%%%%%%%%%%%%%%%%%%%%%
%%%%%%%%%%%%%%%%%%%%%%%%%%%%%%%%%%%%%%%%%%%%%%%%%%%%%%%

\section{Leptogenesis via oscillating Weinberg operator}\label{sec:oscillating}

In the main body of the text, we have assumed the scalar $\phi$ EEV varies smoothly from 0 to $v_\phi$, which should be understood as the ``macroscopic'' behaviour  of $\phi$ during the vacuum transition. At the ``microscopic scale'', the scalar $\phi$ may oscillate, which leads to fluctuations in addition to the ``macroscopic'' behaviour. In this Appendix, through the inclusion of  the oscillation effect, we provide additional details of the  energy transfer from the vacuum to the plasma and discuss the validity of setting the upper bound of the energy transfer to approximately the plasma temperature.

We begin with applying the EEV profile of an oscillating field as shown in Refs.~\cite{Drewes:2010pf, Cheung:2015iqa} in the quasiparticle approximation, and write the EEV profile in the form
\bq
\langle \phi(t) \rangle = [ \langle \phi(t) \rangle|_{t=t_0} \cos (m_\phi (t-t_0)) + m^{-1}_\phi \partial_t \langle \phi(t) \rangle|_{t=t_0} \sin (m_\phi (t-t_0)) ] e^{-\gamma_\phi (t-t_0)} 
\nq
for $t \geqslant t_0$. 
The solution to the  above equation describe the oscillation of $\phi$ near the minimum of its potential.
The plasma frequency, $m_\phi$, is the thermal mass which is found by evaluating the dispersion relation of this 
scalar near zero momentum. As in \cite{Cheung:2015iqa}, we assume a narrow damping rate $\gamma_\phi$ with $\gamma_\phi \ll m_\phi$ which corresponds to $\phi$ scattering with the thermal bath and its decay. 
We do not consider spatial variation in order to limit the complexity of the discussion. 
This profile is obtained by assuming the initial deviations $\langle \phi(t) \rangle - v_\phi$ and $\partial_t \langle \phi(t) \rangle$ after $t_0$ are small \cite{Drewes:2010pf}, such that the mass terms dominates the variation of the scalar EEV. Therefore, it is only valid if the profile of $\langle \phi \rangle$ varies slowly in a certain regime, e.g., towards the end of the phase transition, as well as the end of reheating processes \footnote{Our calculation has assumed thermal distribution for the lepton and Higgs. Thus, by applying the mechanism to reheating processes, one has assumed leptogenesis take places at the very end of reheating when bath has been produced thermally. One could also consider a scenario in which, during reheating, the Higgs or lepton are not fully thermalised. In this case, if the spatial isotropy is still satisfied, we could simply introduce new parameters $\xi_{\ell}=n_\ell/ n_{\ell}^{\rm eq}$ and $\xi_{H}=n_H/ n_{H}^{\rm eq}$, representing the ratio of the true number density to its thermalised number density \cite{DiBari:2016guw}. In this way, the generated lepton asymmetry could be modified by timing a factor $\xi_\ell^2 \xi_H^2$. To generate energy baryon asymmetry, the energy scale should be enhanced by a factor $\xi_\ell \xi_H$. }. For the entire period of phase transition, quartic terms are crucial and affect the running of the scalar mass and thus we cannot apply this approximation. Therefore, the lepton asymmetry generated from this EEV profile should not be understood as the entire  lepton asymmetry. 

Following the oscillating EEV profile, the effective profile of $\lambda$ may be written as 
\bq
\lambda(t)= \lambda + [(\lambda_0-\lambda) \cos (m_\phi (t-t_0)) + m^{-1}_\phi \dot{\lambda}_0 \sin (m_\phi (t-t_0))] e^{-\gamma_\phi  (t-t_0)} \,,
\nq
where $\lambda_0$ and $\dot{\lambda}_0$ are abbreviations of $\lambda(t)$ and $\partial_{t}\lambda(t)$ at $t=t_0$, respectively.  
We apply this profile to demonstrate how the energy transfer from the false vacuum to the plasma is related to the oscillation frequency, $m_\phi$. 

We remind the reader  that the profile $\lambda(t)$ is only valid at the end of the bubble wall with $t\geqslant t_0$. This implies that only lepton asymmetry for $x'$ from $x'_0$ to infinity can be calculated. Thus, the limits of the integration should be replaced, $\int_{t_0}^{\infty}dt_1\int_{t_0}^{\infty}dt_2 = 2 \int_{0}^{\infty}d y \int_{t_0+y/2}^{\infty}dt$. The integration $\int_{t_0+y/2}^{\infty}dt \text{Im}[\lambda^*(t_1)\lambda(t_2)]$ is simplified to
\bq \label{eq:integrate_t}
\int_{t_0+\frac{y}{2}}^{\infty}\!\! dt \text{Im}[\lambda^*(t_1)\lambda(t_2)] 
&=& \frac{\text{Im}[\lambda_0\lambda^*] m_\phi }{m^2_\phi + \gamma_\phi^2} \underbrace{ \left\{ \sin(m_\phi y) e^{-\gamma_\phi y} + \frac{\gamma_\phi}{m_\phi} \left[1 - \cos(m_\phi y) e^{-\gamma_\phi y} \right]\right\} }_{g_1(y)} \nonumber\\
&+& \frac{\text{Im}[\dot{\lambda}_0\lambda^*] }{m^2_\phi + \gamma_\phi^2} \underbrace{ \left\{ 1 - \cos(m_\phi y) e^{-\gamma_\phi y} - \frac{\gamma_\phi}{m_\phi} \sin(m_\phi y) e^{-\gamma_\phi y} \right\} }_{g_2(y)} \nonumber\\
&+& \frac{\text{Im}[ (\lambda_0 - \lambda) \dot{\lambda}^*_0]}{2 m_\phi \gamma_{\phi}} \underbrace{ \sin(m_\phi y) e^{-\gamma_\phi y} }_{g_3(y)} \,.
\nq
In the limit $y \ll m^{-1}_\phi$, $g_{1,3}(y)\simeq m_\phi y$, $g_{2}(y)\simeq 0$, and we recover the result of \equaref{eq:int_x}. 

The lepton asymmetry in this case is replaced by
\bq 
\Delta n_\ell &=& - \frac{12}{v_H^4} \text{Im}[m_{\nu,0}m_\nu^*] \frac{ m_\phi }{m^2_\phi + \gamma_\phi^2} \int d^4r 
g_1(y) \,\mathcal{M} \nonumber\\
&& - \frac{12}{v_H^4} \text{Im}[\dot{m}_{\nu,0}m_\nu^*] \frac{1}{m^2_\phi + \gamma_\phi^2}  \int d^4r 
g_2(y) \,\mathcal{M} \nonumber\\
&& - \frac{12}{v_H^4} \text{Im}[ (m_{\nu,0} - m_\nu) \dot{m}^*_{\nu,0}] \frac{1}{2 m_\phi \gamma_{\phi}}  \int d^4r \, 
g_3(y) \,\mathcal{M} \,,
\label{eq:lepton_asymmetry_f1_app}
\nq
where $\int d^4 r = \int d^3\mathbf{r} \times 2 \int_0^{\infty} dy$. 
Here, the effective neutrino mass matrices $m_{\nu,0} = \lambda_0 v_H^2/\Lambda$, $\dot{m}_{\nu,0} = \dot{\lambda}_0 v_H^2/\Lambda$ are understood. 
Compared with $\Delta n_\ell$ in \equaref{eq:lepton_asymmetry_f1}, the main difference in \equaref{eq:lepton_asymmetry_f1_app} in addition to the neutrino mass combinations, is the integrand $y\mathcal{M}$ replaced by $g_{1,2,3}(y)\mathcal{M}$, where $g_{1,2,3}(y)$ have been defined in \equaref{eq:integrate_t}. 
We follow the same procedure as applied in \secref{sec:thermaleff} to integrate out $d^3\mathbf{r}$ and $d^3\mathbf{k}'$ and then arrive at the integration $2\int_{0}^{+\infty} dy g_{1,2,3}(y)M$. 
The integral $\int_{0}^{+\infty} dy \cos(my) e^{-\gamma_\phi y}M$ and $\int_{0}^{+\infty} dy \sin(my) e^{-\gamma_\phi y}M$ appear to take a common factor, which can be rewritten in the following form 
\bq
\frac{K \sinh  (\beta K/2)}{[(K-m)^2+\tilde{\gm}^2][(K+m)^2+\tilde{\gm}^2]} = \frac{1}{4m } \left( \frac{\sinh (\beta K/2)}{[(m-K)^2+\tilde{\gm}^2]} + \frac{\sinh  (\beta (-K)/2)}{[(m+K)^2+\tilde{\gm}^2]}  \right),
\nq
where $K$ represents any of $K_{\eta_2 \eta_3 \eta_4}$ and $\tilde{\gm} = \gm + \gamma_\phi$ is  the total summed damping rate of the  leptons, Higgses and the oscillating scalar $\phi(t)$. 
By defining 
\bq
\Delta_{\gm}(m-K) = \frac{\gm}{(m-K)^2 + \gm^2},
\nq
this common factor is reexpressed as 
\bq
\frac{1}{4m \tilde{\gm}} \sum_{\eta_1 = \pm 1 } \sinh(\beta \eta_1 K/2) \Delta_{\tilde{\gm}}(m+\eta_1 K) \,.
\nq
Taking into account of the sign $\eta_1$, we can expend $K_{\eta_2 \eta_2 \eta_3}$ to 
\bq
K_{\eta_1 \eta_2 \eta_3 \eta_4} = \eta_1 \omega_{\bfk} + \eta_2 \omega_{\bfk'} + \eta_3 \omega_{\bfq} + \eta_4 \omega_{\bfq'} \,.
\nq
With this treatment, the integrals $2\int_{0}^{+\infty} dy g_{1,2,3}(y)M$ are simplified into forms
\bq 
\int_{0}^{+\infty} dy g_1(y)M &\approx& \int_{0}^{+\infty} dy g_3(y)M \nonumber\\
&\approx& \sum_{\eta_1,\eta_2,\eta_3,\eta_4=\pm1} \frac{ [1- \eta_1 \eta_2\hat{\bfk}\cdot \hat{\bfk}'] }{128 \omega_{\bfq} \omega_{\bfq'} A } 
\sinh(\beta K_{\eta_1 \eta_2 \eta_3 \eta_4}/2) \Delta_{\tilde{\gm}} (m-K_{\eta_1 \eta_2 \eta_3 \eta_4}) \,, \nonumber\\
\int_{0}^{+\infty} dy g_2(y)M &\approx& \lim_{\epsilon \to 0} \sum_{\eta_1,\eta_2,\eta_3,\eta_4=\pm1} \frac{ [1- \eta_1 \eta_2\hat{\bfk}\cdot \hat{\bfk}'] }{256 \omega_{\bfq} \omega_{\bfq'} A } \nonumber\\
&&
\times \left\{ \frac{K_{\eta_1 \eta_2 \eta_3 \eta_4}^2 - \epsilon^2 + \gm^2}{\epsilon \gm} 
\sinh(\beta K_{\eta_1 \eta_2 \eta_3 \eta_4}/2) \Delta_{\gm} (\epsilon-K_{\eta_1 \eta_2 \eta_3 \eta_4}) \right.\nonumber\\
&&\left.-\frac{K_{\eta_1 \eta_2 \eta_3 \eta_4}^2 - m_\phi^2 + \tilde{\gm}^2}{m_\phi \tilde{\gm}}
\sinh(\beta K_{\eta_1 \eta_2 \eta_3 \eta_4}/2) \Delta_{\tilde{\gm}} (m_\phi-K_{\eta_1 \eta_2 \eta_3 \eta_4}) \right\} \,,
\label{eq:LeptonAsymm3}
\nq
where
\bq
A = \cosh(\omega_{\bfk} \beta/2) \cosh(\omega_{\bfk'} \beta/2) \sinh(\omega_{\bfq} \beta/2) \sinh(\omega_{\bfq'} \beta/2) \,.
\nq
Here, we have ignored terms of higher orders of $\gamma_\phi$, and 
\bq
\frac{K\sinh(\beta K/2)}{K^2+\gm^2} &=& \lim_{\epsilon \to 0} \frac{(K^2-\epsilon^2+\gm^2)K \sinh  (\beta K/2)}{[(K-\epsilon)^2+\gm^2][(K+\epsilon)^2+\gm^2]} \nonumber\\
&=& \lim_{\epsilon \to 0} 
 \sum_{\eta_1 = \pm 1 } \frac{K^2-\epsilon^2+\gm^2}{4\epsilon \tilde{\gm}} \sinh(\beta \eta_1 K/2) \Delta_{\tilde{\gm}}(\epsilon+\eta_1 K) \,,
\nq

Again in the limit $y \ll m^{-1}_\phi$, we recover \equaref{eq:inteM} from the integrals involving $g_1(t)$ and $g_3(t)$, while the integral involving $g_2(t)$ vanishes. Thus, this result is compatible with that in the main text. 

We would like to explore the case of vanishing damping rates as this will illustrate a limit (albeit unphysical)
of energy transfer between the scalar field and the thermal plasma.
Although this case is unphysical, it is instructive to start from this limit as it  shows similarities and differences of our work with the classical QFT, where incoming and outing particles are treated as free particles. By setting $\gm, \tilde{\gm}\to 0$, we arrive at
\bq
\lim_{\epsilon \to 0} \Delta_{\gm}(\epsilon-K_{\eta_1 \eta_2 \eta_3 \eta_4}) &=&\lim_{\epsilon \to 0} \pi \delta (\epsilon - K_{\eta_1 \eta_2 \eta_3 \eta_4}) = \pi \delta (K_{\eta_1 \eta_2 \eta_3 \eta_4}) \,, \nonumber\\
\Delta_{\tilde{\gm}}(m_\phi - K_{\eta_1 \eta_2 \eta_3 \eta_4}) &=& \pi \delta (m_\phi-K_{\eta_1 \eta_2 \eta_3 \eta_4}) \,,
\label{eq:delta}
\nq
respectively. 
The first $\delta$ function implies energy conservation during the scattering of leptons and the Higgses via Weinberg operator.
The second $\delta$ function leads to $K_{\eta_1 \eta_2 \eta_3 \eta_4} = m_\phi$. This  shows the energy transfer between $\phi$ and thermal bath particles (i.e., leptons and Higgses) is $m_\phi$. 
Although, consideration of  zero limit of the damping rates may be helpful for understanding the energy transfer, it conceals some crucial contributions in our mechanism of leptogenesis: 
\begin{itemize}
\item
By setting the lepton and Higgs damping rates ($\gamma_\ell$, $\gamma_H$) to zero, all off-shell processes related to these particles, e.g., transition emission from a lepton after it is produced by the Weinberg operator, are forbidden. The energy transfer between the scalar and the thermal bath has to be fixed at $0$ or $m_\phi$, refer to the first and second $\delta$ functions in \equaref{eq:delta}, respectively. 

\item
By setting the damping rate of the scalar EEV ($\gamma_\phi$) to zero, the scalar has a stable oscillating profile with no damping. Therefore, processes of the scalar releasing energy to the plasma take place in half of one period and the reverse processes take place in the other half period with the same strength. As a consequence, a positive lepton asymmetry is generated in one half period, while the same amount of negative lepton asymmetry is generated in the other half period. Therefore, the lepton asymmetry oscillates (does not converge) with time. We note that this divergent behaviour is also reflected in \equaref{eq:inteM} in the zero-width limit. 

\end{itemize}
For illustration, we calculate the integrals $2\int_{0}^{+\infty} dy g_{1,2,3}(y)M$ in the limit of zero damping widths. It is straightforward to obtain  
\bq 
\int_{0}^{+\infty} dy g_{1,3}(y)M &\to& \int_{0}^{+\infty} dy \sin(m_\phi y)M|_{\gamma = 0} \nonumber\\
\int_{0}^{+\infty} dy g_2(y)M &\to& 0 \,,
%\label{eq:LeptonAsymm3}
\nq
from the definitions of $g_{1,2,3}(y)$ in Eq.~\eqref{eq:integrate_t}. 
Then, following Eq.~\eqref{eq:inteM}, we arrive at 
\bq 
&&\int_{0}^{Y} dy \sin(m_\phi y)M|_{\gamma = 0}  \nonumber\\
&&=  \int_{0}^{Y} dy \sin(m_\phi y) \frac{\text{Im} \{ [c(\omega_{\bfk}y^-)c(\omega_{\bfk'}y^-)+\hat{\bfk}\cdot \hat{\bfk}' s(\omega_{\bfk} y^-) s(\omega_{\bfk'} y^-)] c(\omega_{\bfq}y^-)c(\omega_{\bfq'}y^-) \}}
{8\omega_{\bfq} \omega_{\bfq'} ch(\omega_{\bfk} \beta/2) ch(\omega_{\bfk'} \beta/2) sh(\omega_{\bfq} \beta/2) sh(\omega_{\bfq'} \beta/2) }  \nonumber\\
&&= \sum_{\eta_1, \eta_2,\eta_3,\eta_4=\pm1} \frac{ [1- \eta_1 \eta_2\hat{\bfk}\cdot \hat{\bfk}'] \sin[(K_{\eta_1 \eta_2 \eta_3 \eta_4} -m_\phi) Y] \, sh(\beta K_{\eta_2\eta_3\eta_4} /2)}{16 \omega_{\bfq} \omega_{\bfq'} (K_{\eta_1 \eta_2\eta_3\eta_4} -m_\phi) \, ch(\omega_{\bfk} \beta/2) ch(\omega_{\bfk'} \beta/2) sh(\omega_{\bfq} \beta/2) sh(\omega_{\bfq'} \beta/2)}\,. \nonumber\\
%\label{eq:LeptonAsymm3}
\nq
This result oscillates with time difference $Y$, as we already mentioned above. It does not converge when $Y \to \infty$ except including the damping effect. In the limit $Y \to 0$, we cover a momentum conversation $K_{\eta_1 \eta_2\eta_3\eta_4} -m_\phi$ with the help of $\delta(x) = \displaystyle \frac{1}{\pi}\lim_{Y\to 0} \frac{\sin(xY)}{x}$.
We emphasise that  the zero width limit is an unphysical if we do not set $\lambda=\lambda_0=0$ (otherwise the coupling to leptons and Higgses would cause $\phi$ to have a non-zero width). Consistently taking these limits together causes the integral \equaref{eq:integrate_t} to be zero and therefore not divergent.

To summarise, we introduce a non-zero damping rates of the lepton and Higgs to parametrise off-shell effects related to these particles. As the  leptons or Higgses may transfer their energy to other degrees of freedom in the thermal bath, the energy released to the plasma ($K_{\eta_1 \eta_2 \eta_3 \eta_4} $)  does not need to be $m_\phi$, but be in a range around $m_\phi$. A naive estimation of the scalar mass is that it is in the same order of the temperature $T$. Thus, we set an upper bound for the energy transfer around $T$. 
We note that this effective treatment is adopted in order to remain agnostic about details of the scalar, $\phi$,  such as its mass and its precise microphysical interactions with the leptons and Higgs.

We include a non-zero damping rate for the scalar EEV to drive the EEV in a definite direction, i.e., $\langle \phi(t) \rangle$ varying from $\langle \phi(t) \rangle|_{t=t_0}$ to $0$, as well as $\lambda(t)$ varying from $\lambda_0$ to $\lambda$, as well as to obtain a net energy transfer from the vacuum to the plasma. While the oscillating damping EEV profile does not apply to the whole period of phase transition, an alternative effective treatment is to consider only the ``macroscopic''  behaviour of $\lambda(t)$: running definitely from initial value $\lambda^{0}$ to final value $\lambda$.

\bibliographystyle{JHEP}
\bibliography{ref}

\end{document}